\documentclass{statsoc}

\RequirePackage{amsmath}
\RequirePackage{natbib}
\RequirePackage{hyperref} 
\RequirePackage{amssymb}
\RequirePackage{graphicx}
\RequirePackage{bbm}
\RequirePackage{subfigure}
\RequirePackage{xcolor}
\RequirePackage{fullpage}
\usepackage[a4paper]{geometry}
\usepackage{graphicx}
\usepackage[textwidth=8em,textsize=small]{todonotes}
\usepackage{amsmath}
\usepackage{natbib}

\title{Network Hawkes Process Models for Exploring \\Latent Hierarchy in Social Animal
Interactions}
\author[Ward {\it et al.}]{Owen G.\ Ward*, Jing Wu*,  Tian Zheng}
\address{Department of Statistics, Columbia University, New York, NY.}
\author{Anna L. Smith}
\address{Department of Statistics, University of Kentucky.}
\author{James P. Curley}
\address{Department of Psychology, University of Texas at Austin}

\begin{document}
\let\thefootnote\relax\footnote{*These authors contributed equally. Correspondence should be addressed to: Tian Zheng, 1255 Amsterdam Avenue, MC 4690, New York, NY 10027; tzheng@stat.columbia.edu. }

\begin{abstract}
Group-based social dominance hierarchies are of essential interest in understanding social structure \citep{dedeo2021equality}. Recent animal behavior research studies can record aggressive interactions observed over time. Models that can explore the underlying hierarchy from the observed temporal dynamics in behaviors are therefore crucial. Traditional ranking methods aggregate interactions across time into win/loss counts, equalizing dynamic interactions with the underlying hierarchy. Although these models have gleaned important behavioral insights from such data, they are limited in addressing many important questions that remain unresolved. 
In this paper, we take advantage of the observed interactions' timestamps,
proposing a series of network point process models with latent ranks. 
We carefully design these models to incorporate important theories on animal behavior 
that account for dynamic patterns observed in the interaction data, 
including the winner effect, bursting and pair-flip phenomena.
Through iteratively constructing and evaluating these models
we arrive at the final cohort Markov-Modulated Hawkes process (C-MMHP), which best 
characterizes all aforementioned patterns observed in interaction data.
 
%
As such, inference on our model components can be readily interpreted in terms of theories on animal behaviors.
The probabilistic nature of our model
allows us to estimate the uncertainty in our ranking.
In particular, our model is able to provide insights into
the distribution of power within the hierarchy which forms
and the strength of the 
established hierarchy.
We compare all models using simulated and real data.
Using statistically developed
diagnostic perspectives, we demonstrate
that the C-MMHP model outperforms other methods, 
capturing relevant latent ranking structures that lead to meaningful predictions for real data.
\end{abstract}

\section{Introduction}

In this paper we consider the problem of
providing a general model-based framework for exploring the unobserved social hierarchy among a group of mice through their observed repeated aggressive interactions. We do this using data from the study conducted by \cite{williamson2016temporal}, in order to address unsolved questions in that work.  In particular,
describing the dominance structure
behind such interactions well is a difficult task,
and existing methods cannot adequately
capture all possible dynamics, quantify uncertainty in the ranking,
or provide insights into how a power hierarchy is formed or the distribution
of that hierarchy.

There is no existing method which can describe how the observed interactions
are generated from the underlying social hierarchy.
Similarly, how mice are able to recognize their social status relative to other mice and how this recognition facilitates hierarchy formation and maintenance remains an unanswered question.
The temporal dynamics in these interactions are driven by the need of the mice
to explore, recognize, maintain, and exploit their positions in such a hierarchy, through mechanisms that are not fully understood.
Section \ref{sec_latent:related} presents an overview of existing well-known methods for dominance ranking and their properties. These existing methods,
which generally utilise aggregate data,
suffer several
common issues, including the inability to rigorously evaluate the estimated ranking and the 
inability to deal with the temporal component of 
these interactions,
which is likely influenced by the animals' gains of social information about their group's structure \citep{hobson2021aggression}.
Specifying statistical generative models therefore 
provides a natural way to characterize the structure of these social 
groups more generally.
One focus of the models we develop here is the
ability to capture the latent stable dominance hierarchy via modeling the temporal and network dynamics of
these social interactions.
In Section \ref{sec_latent:models}, we take advantage of the
timestamps of these 
interactions and propose three network point process models: the
cohort Hawkes process model (C-HP), the cohort degree-corrected 
Hawkes process model (C-DCHP) and the cohort Markov-modulated Hawkes 
process model (C-MMHP).
We construct these models such that
these point processes 
are a function of a set of latent rank variables. These
latent rank variables are a powerful feature of our models,
allowing us to incorporate various known
traits
of animal behaviors in a social hierarchy into our model. We develop these models in a 
Bayesian framework to capture uncertainty estimates and
to better model pairs which contain few interactions.
We iteratively develop
each model from the previous to better account for dynamics seen in animal 
data. This results in our final
Cohort Markov-Modulated Hawkes Process (C-MMHP)
model. In Section \ref{sec_latent:results}, these 
models are compared,
using simulated and real data,
to existing methods for understanding animal
dominance ranking,
highlighting how different methods capture different behavior components,
leading to different rank estimates.
We illustrate 
that our final model is flexible and adequately captures dynamics driven by the group's inherent
dominance hierarchy by showing results on rank inference, prediction performance and 
residual analysis.
These point process models, therefore, provide a new utility for future research that can lead to better understandings of the 
dominance hierarchies among animals and be used to generate further research questions.
Section 
\ref{sec_latent:discussion} summarizes this work and discuss future
directions for our proposed model.



\section{Background}
\label{sec_latent:related}
Here we review the literature on social hierarchy for group-living animals. 
Empirical studies of the social hierarchy of animals that live in a group
are generally developed based on the 
observations of dyadic, or pairwise, agonistic interactions. In 
\cite{williamson2016temporal}, the agonistic interactions include fighting, 
chasing and mounting behaviors. We consider all such 
aggressive interactions without differentiating the type,
as is often done in this area \citep{lee2019temporal}.
We denote 
the interactions between $N$ animals as a matrix $W$, where $W_{ij}$ 
is the number of aggressive interactions 
 won
by animal $i$ against animal $j$.  In 
\cite{so2015social} and \cite{williamson2016temporal}, this is also 
called a \textit{win/loss} matrix.

Two approaches are generally considered in the animal behavior
literature to analyse this win/loss matrix \citep{drews1993concept}:
\textit{functional} methods and \textit{structural} methods.
\textit{Functional} methods aim to directly infer
a ranking of animals from this win/loss matrix
by rearranging this matrix in an attempt to best
capture behavioral patterns,
expected in a social hierarchy.
The rank is therefore inferred directly from the observations
recorded in the \textit{win/loss} matrix. If $W_{ij} > W_{ji}$
then functional methods infer that $i$ dominates $j$.
Alternatively, \textit{structural} methods propose
an indirect model-based approach, associating
a latent ranking variable $F_i$ with individual $i$.
If $F_i > F_j$ then these \textit{structural} methods
infer that $i$ dominates $j$.
These latent variables are 
constructed to
satisfy a set of a priori
assumptions, and structural models attempt
to estimate these latent ranks to best align
with the behavior captured in $W$.

An important concept in dominance ranking is \textit{linearity}.
Under a strict linearity assumption, for any three individuals,
$i,j$, and $k$, if $i$ dominates $j$ and $j$ dominates $k$, then 
$i$ is assumed to dominate $k$. In social network research,
this closed triad relationship is also called \textit{transitivity}. 
For \textit{functional} methods, the linearity assumption intuitively 
follows from observational studies of group-living animals. However,
phenomena that violate a strict linearity assumption are often  observed.
In these cases, \textit{functional} methods aim to find a
nearly linear ranking that is most consistent with the observed wins and losses. 
Meanwhile,
in \textit{structural} methods, the linearity assumption is not directly observable
but is incorporated as a property of the latent parameter $F$. 
The goal for \textit{structural} methods is to study the 
model that can mostly reflect the potential formation mechanisms of dominance hierarchy. 

\begin{figure}[ht]
 \centering
 \includegraphics[width=0.8\textwidth]{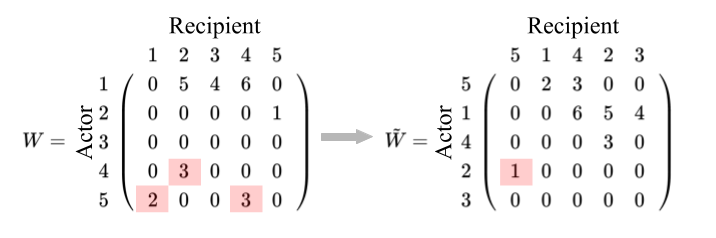}
 \caption{An example of a \textit{win/loss} matrix and the corresponding reordered matrix according to 
 the I\&SI method. The entries shaded in red in the matrix are the \textit{inconsistencies}, where the 
 lower-ranked individual wins more frequently than the higher-ranked individual.}
 \label{fig_latent:isi_matrix}
\end{figure}

One popular functional
model is the I\&SI method of \cite{de1998finding}.
The I\&SI method is a matrix-reordering method that identifies ordinal
rankings of individuals that are most consistent with a linear hierarchy,
by iteratively minimizing two criteria: the number of inconsistencies 
(\textit{I}) and then, conditionally, the total strength of the 
inconsistencies (\textit{SI}) without increasing \textit{I}. 
The number of inconsistencies (\textit{I}) is the number 
of pairs in which the lower-ranked individual wins more
frequently than the higher-ranked individual in a given
\textit{win/loss} matrix, $\tilde{W}$, 

$$
I=\sum_{i>j}\mathbbm{1}_{\{\tilde{W}_{ij}>\tilde{W}_{ji}\}},
$$
where $\mathbbm{1}_{\{\cdot\}}$ is an indicator function. The matrix 
$\tilde{W}$ is generated by reordering the original win/loss matrix 
$W$ according to a ranking of the individuals. The \textit{strength} of a single 
inconsistency is the absolute rank difference of the inconsistent
pair. Then, the total strength of the inconsistencies (\textit{SI}) 
is the sum of strengths of all inconsistencies in $\tilde{W}$, 
$$
SI =
\sum_{i>j}|i-j|\mathbbm{1}_{\{\tilde{W}_{ij}>\tilde{W}_{ji}\}}.
$$
An example is shown in Figure \ref{fig_latent:isi_matrix}. The 
original \textit{win/loss} matrix in the example is $W$, which 
corresponds to $I=3$ and $SI=7$. According to the I\&SI ranking 
method, the matrix is reordered to yield $\tilde{W}$, the rightmost 
matrix in Figure \ref{fig_latent:isi_matrix}, in which $I=1$ and 
$SI=3$. Intuitively, the I\&SI method finds the order of the rankings
that is most consistent with a linear hierarchy. Although such a 
perfect linear hierarchy usually does not exist, I\&SI aims 
to find a ranking where any inconsistencies take place between 
individuals that are close in rank. In other words, the 
I\&SI method is most likely to allow for inconsistent dyads near the diagonal.

This method suffers from the problem that the algorithm is not guaranteed to converge to a unique
optimal solution \citep{vries2000finding}. In particular, when there is a tie in the number of
wins/losses ($W_{ij}=W_{ji}>0$) or an unknown relationship (i.e., where there is little information, 
$W_{ij}=W_{ji}=0$), the result highly depends on the choice of rules for assigning rankings. Another 
reason for the divergence is the method's reliance on only an asymmetric relationship between the number
of wins and losses, instead of the absolute difference. Such a simplified binary dominance measure
ignores important information in the data -- the total number of fights. It is often observed that the 
distribution of dominance power exhibits a high discrepancy \citep{chase2002individual},
such as when
highly ranked animals win a larger number
of fights against intermediately ranked animals than 
these intermediate 
animals win against lowly ranked animals.
This is seen in a \textit{win/loss} matrix with
large variation in the values of $W_{ij}$,
but an ordinal ranking from a binary 
dominance measure is not discriminative enough to demonstrate that. \cite{williamson2016temporal}
provide an analysis of monopolization of the most dominant mouse in each cohort, which suggests the 
necessity for considering a real-valued score instead of the ordinal rank.

Winner-loser models are an important family of structural methods 
that aim to explain the 
formation of linear dominance hierarchy \citep{lindquist2009data}. 
Commonly, the 
models assume an innate power parameter for each individual $i$, 
denoted as $F_i$ 
(some models may assume a time-variant version, i.e. $F_i(t)$) 
\citep{bonabeau1999dominance, dugatkin1997winner, 
hemelrijk2000towards}. Although 
different models have their own specific formulations, 
common components they share 
are: an interaction probability and a dominance probability. Both are
functions of 
innate power. The mathematical formulation of the models will not be 
discussed 
here, but some assumptions used in these model are of interest. 
One essential 
idea is the \textit{winner effect},
the phenomenon in which an animal
that has 
experienced previous wins will continue to win future aggressive 
interactions with 
increased probability. Although the extent and 
effectiveness of these 
winner effects 
remains unclear, evidence from experiments show that they exist 
and vary in 
different groups and species 
\citep{dugatkin1997winner, dugatkin2003group, hsu1999winner}. 
Experimental evidence also shows patterns that are 
challenging to capture through winner-loser models, such 
as \textit{bursting} 
and \textit{pair-flips}. Bursting means that higher-rank animals 
often exhibit 
successive fighting of lower-rank ones in an extended period of time.
Pair-flips 
describe the situation when a pair of animals exchange the direction of their 
aggressive acts before a stable dominance relationship is established.
A potential model for learning a latent hierarchy 
should therefore be able to incorporate these characteristics.
In Section~\ref{sec_latent:results} we describe several
existing structural and functional models 
in detail,
which we
use for comparison with our proposed methods.

\subsection{Issues with conventional approaches}

In summary, although the current methods have led to important insights on social structures among animals
\citep{so2015social, williamson2016temporal,hobson2021aggression},
they suffer from several issues that prevent them from being of utility for today's increasingly available data with temporal information. 

For functional methods, the use of an ordinal ranking alone may not be informative enough to describe the unequal distribution of dominance power which is commonly observed.
The observed win/loss matrices are a noisy realisation of the true underlying
dominance ranking among animals. Algorithms that attempt to directly derive the ranking by 
rearranging $W$ can therefore be unstable and unable to account for even small
deviations from expected behavior. In addition, uncertainty
measures are not available for the dominance ranking produced.

Structural models for such data provide much potential to understand the underlying drivers of animal  behavior. However, methods to evaluate these models have been limited,
as seen in \cite{lindquist2009data}. These models are generally not built under the statistical framework of generative models. 
%
%
They fall short of providing a probabilistic connection between the observed social interaction timestamps and the underlying dominance hierarchies, and as such are unsuited to rationalize noisy patterns against the subject-matter assumptions of these models.
As a result, a more systematic solution is needed for modeling the temporal dynamics of the dominance hierarchy, instead of relying on empirical scores. The timestamps of interactions among a group contain information about  how the particular hierarchy formation and social information of this hierarchy are associated with social interaction patterns over time. This framework can provide better insights into the questions 
of \cite{williamson2016temporal} such as agnostic interactions in uncommon 
directions.
The utilisation of a probabilistic generative model for latent ranking further 
provides the potential to rigorously assess model fit and help formalise scientific 
hypotheses.
In the next section we will develop 
generative point process network models for
this data, which we then compare with these
existing methods in Section~\ref{sec_latent:results}.


\section{Latent ranking structured network point process models}

In this section, we propose a 
series
of probabilistic generative models to address common issues with conventional approaches, as discussed in the previous section. Inspired by theories on social hierarchy among group-living animals, there are 
various properties that we want to
take into account when constructing these models: 
inconsistencies lying between the interactions and rankings, the
time-evolving nature of the interaction dynamic, the winner effect, bursting and 
pair-flips phenomenons. For modeling the observed timestamps of social interactions, our models address three contributing mechanisms: the unobserved social hierarchy, the developmental influence from the animals' social information about the social hierarchy, and temporal dependence on historical interactions.


\label{sec_latent:models}

We first introduce the
required notation of point process models and network data, 
leading to 
a model for network point processes 
(Section~\ref{sec_latent:back_netpp}).
We then propose three such network point process models based on latent (structural) rankings (Section~\ref{sec_latent:model_netpp}).
We motivate the development of each of these models in turn by examining the properties each model fails to capture in one cohort of
mice interaction data (described in Section~\ref{sec_latent:real_data}),
using the
inference procedure described in 
Section~\ref{sec_latent:model_fitting}.

\subsection{Network point process models for animal interactions}
\label{sec_latent:back_netpp}

Animal aggressive interaction data is essentially network data,
where the senders are the winners of the fights and the receivers are the losers. To consider the necessary information lying in the timestamps of interactions, we introduce point process models
on networks. In this section, we start with notation and a 
discussion on point processes in general (i.e.,
for non-network data), with a focus on the Hawkes process.
We then introduce the notation for network event 
arrival data and network point process models. 

\paragraph{Point process models.} Consider event arrival time 
data that consists of all event history up to a 
{\em final-observation} time $T$:
$\mathcal{H}(T) = \{t_m\}_{m=0}^M$,
where $t_0=0$, $t_M=T$, and $M$ is the total number of events.
An equivalent representation of this event history $\mathcal{H}(T)$
is via a {\em counting process}, $N(t)$, where $N(t)$ is a
right-continuous function that records the number of events
observed during the interval $(0,t]$.
The associated stochastic property is usually specified
by its conditional intensity function 
$\lambda(t|\mathcal{H}(t))$ 
at any time $t\in(0,T]$,
conditioning on current history $\mathcal{H}(t)$,
$$\lambda(t|\mathcal{H}(t))
=
\lim_{\Delta t\to 0} \frac{Pr(N(t+\Delta t)-N(t)=1|\mathcal{H}(t))}{\Delta t}.$$
This is the instantaneous expected rate of events
occurring around a time $t$ given the history.
Inference on the intensity function 
is conducted by evaluating the likelihood function for a sequence of events up to 
time $T$, $\mathcal{H}(T)$, which can be expressed as \citep{daley2003introduction}
\begin{equation}
\label{eq_latent:likelihood}
\prod_{m=1}^M \lambda(t_m|\mathcal{H}(t_m))\exp\Big\{-\int_{0}^{T} \lambda(s|\mathcal{H}(s)) ds\Big\}.
\end{equation}

A {\em Hawkes process} \citep{hawkes1971spectra} is a linear self-exciting 
process that can explain bursty patterns in event dynamics. For a univariate 
model, the intensity function with exponential triggering function is defined as
\begin{equation}
 \lambda(t) = \lambda_1 + \sum_{t_m<t}\alpha e^{-\beta(t-t_m)},
 \label{eqn_hawkes}
\end{equation}
where $\lambda_1 > 0$ specifies the baseline intensity, $\alpha>0$ 
calibrates 
the instantaneous boost to the event intensity at each arrival of an 
event, and 
$\beta>0$ controls the decay of past events' influence over time.


\paragraph{Network point process models.} 
Consider a network 
consisting of a fixed set of $N$ nodes, $V=\{1,2,...,N\}$. For each 
directed 
pair
of nodes $(i,j)$, the observations of interactions (fights) between 
them up to 
terminal time $T$ includes the sender (winner) $i$, 
the receiver (loser) $j$ and
a sequence of event times 
$\mathcal{H}^{i,j}(T):= \{t_m^{i, 
j}\}_{m=0}^{M^{i,j}}$.
 Hence, a network Hawkes process model has a conditional intensity function
for each pair $(i,j)$ at
time $t$ given by $\lambda^{i,j}(t|\mathcal{H}^{i,j}(t))$.
The likelihood of the interactions on the whole network is then
$$\prod_{i=1}^N\prod_{j\neq i}^N\prod_{m=1}^{M^{i,j}} 
\lambda^{i,j}(t^{i,j}_m|
\mathcal{H}^{i,j}(t^{i,j}_m))\exp\Big\{-\int_{0}^{T} 
\lambda^{i,j}(s|\mathcal{H}^{i,j}(s)) ds\Big\}.
$$

\subsection{Latent ranking structured models for network point processes}
\label{sec_latent:model_netpp}

Motivated first by the \textit{winner effect} reviewed in Section 
\ref{sec_latent:related},
we model the conditional intensity of 
directed winning
interactions between a given 
node
pair as a function of their event (winning) history. 
Although experimental observations cannot explicitly verify
the 
extent or persistence of influence of historical events, the intensity formulation 
in the 
Hawkes process (\ref{eqn_hawkes})
can help us model this \textit{winner effect} flexibly. In a 
Hawkes process, $\alpha$ describes the extent to which previous wins 
influence the tendency to engage in a new fight. $\beta$ represents the 
persistence -- how fast this effect decays over time. A large $\beta$ means 
that the winner effect decays quickly and only the most recent wins 
influence the tendency to engage in aggressive interactions at the present 
time.

For a directed pair $(i,j)$, the Hawkes process intensity is
$$
\lambda^{i,j}(t)=\lambda^{i,j}_1+\alpha^{i,j}\sum_k\exp{(-\beta^{i,j}(t-t_
k^{i,j}))},
$$
where $\lambda^{i,j}_1, \alpha^{i,j}$ and $\beta^{i,j}$ are pair-wise 
parameters in the Hawkes process. For all pairs, we will introduce structure in these pair-wise parameters below by assuming a latent rank variable, $f_i\in[0,1],\ i=1,2,...,N$.
This is similar to the latent characteristic concept used in the 
winner-loser models \citep{lindquist2009data} and
the latent rank in the aggregate-ranking model \citep{de2018physical} 
reviewed in Section \ref{sec_latent:related_structural}.
The latent rank variable essentially embeds each
individual in a one-dimensional unobserved ranking space.  We constrain the pair-wise intensity
function by bounding the latent rank in order to avoid issues with model 
identifiability. This latent rank variable is a powerful
feature of our model as it allows us to incorporate various model assumptions 
on how the latent dominance hierarchy and historical events
(i.e., social information) influence social interaction dynamics by specifying particular forms of the parameters $\lambda^{i,j}_1, \alpha^{i,j}$ and $\beta^{i,j}$ in the intensity function, as we will discuss in Section 
\ref{sec_latent:model_1}, \ref{sec_latent:model_2} and 
\ref{sec_latent:model_3}.

\subsubsection{Cohort Hawkes Process (C-HP) Model}
\label{sec_latent:model_1}
In the first model, we assume a baseline intensity, $\lambda_1$,
and that the rate of decay for historical events, $\beta$,
is constant across pairs. We structure the impact of historical
events on each pair as a function of the pair's latent ranks
$f_i,f_j$ and parameters $\mathbf{\eta}$, i.e. 
$\alpha^{i,j}:=g_\mathbf{\eta}(f_i,f_j)$. Inspired by
the inconsistency and strength of inconsistency concepts
in the I\&SI method, we expect that the function
$g_\mathbf{\eta}(f_i,f_j)$ satisfies the following:
(1) $g_\mathbf{\eta}(f_i,f_j)>g_\mathbf{\eta}(f_j,f_i)$
when $f_i>f_j$; (2) $g_\mathbf{\eta}(f_i,f_j)$ is a
decreasing function of $|f_i-f_j|$ when $f_i-f_j<0$. Hence, we consider,
$$
g_\mathbf{\eta}(f_i,f_j):=\eta_1f_if_j\exp{(-\eta_2|f_i-f_j|)}\mbox{logistic}(\eta_3(f_i-f_j)),
$$ 
where $\eta:=(\eta_1,\eta_2,\eta_3)$. 

Each 
component here is motivated by empirical evidence of observed behaviors seen
in animal data. In particular, we describe each term in this
excitation function in detail:
\begin{itemize}
    \item $\eta_1 f_i f_j$ encodes the
    overall tendency of animals to continue fighting as
    a function of the product of their latent 
    rankings. This results in pairs of higher
    ranked individuals being more likely to 
    continue fighting. For higher ranked animals
    there is added incentive to clearly establish their
    dominance over similarly ranked animals, hence more
    repeated interactions. Meanwhile, for a pair of
    lower ranked animals, there is often less
    incentive to move from (say) the lowest rank of the
    hierarchy to the second lowest.
    This behavior is observed empirically in mice, with
    the higher ranked animals being
    most active throughout.
    Modeling the
    excitation parameter with the product of latent
    rankings allows us to capture this variation 
    across pairs.
    \item $ \exp\left( -\eta_2 |f_i-f_j|\right)$ captures
    important properties found in existing methods within
    our model. This component gives
    an excitation parameter of the Hawkes process
    that
    is a decreasing function of $|f_i-f_j|$ when $f_i < f_j$.
    With this condition, this means that a weaker animal
    is less likely to continue 
    winning fights against
    a stronger 
    animal, with the likelihood of these events 
    decreasing as the discrepancy between their latent 
    rankings increases. This is motivated by the strength of
    inconsistencies component of the I\&SI method and agrees with
    observed behavior.
    \item $\mbox{logistic}\left( \eta_3 (f_i -f_j)\right)$
    ensures that $g_{\eta}(f_i, f_j) > g_{\eta}(f_j,f_i)$
    when $f_i>f_j$. This condition means that 
    wins
    are more often from a dominant animal
    to a submissive one. Again, this aligns with
    the inconsistency concept of the I\&SI method while being a 
    property observed in real data.
    
\end{itemize}

\begin{figure}[ht]
 \subfigure
 {
 \includegraphics[width=0.53\linewidth,height=0.46\linewidth]{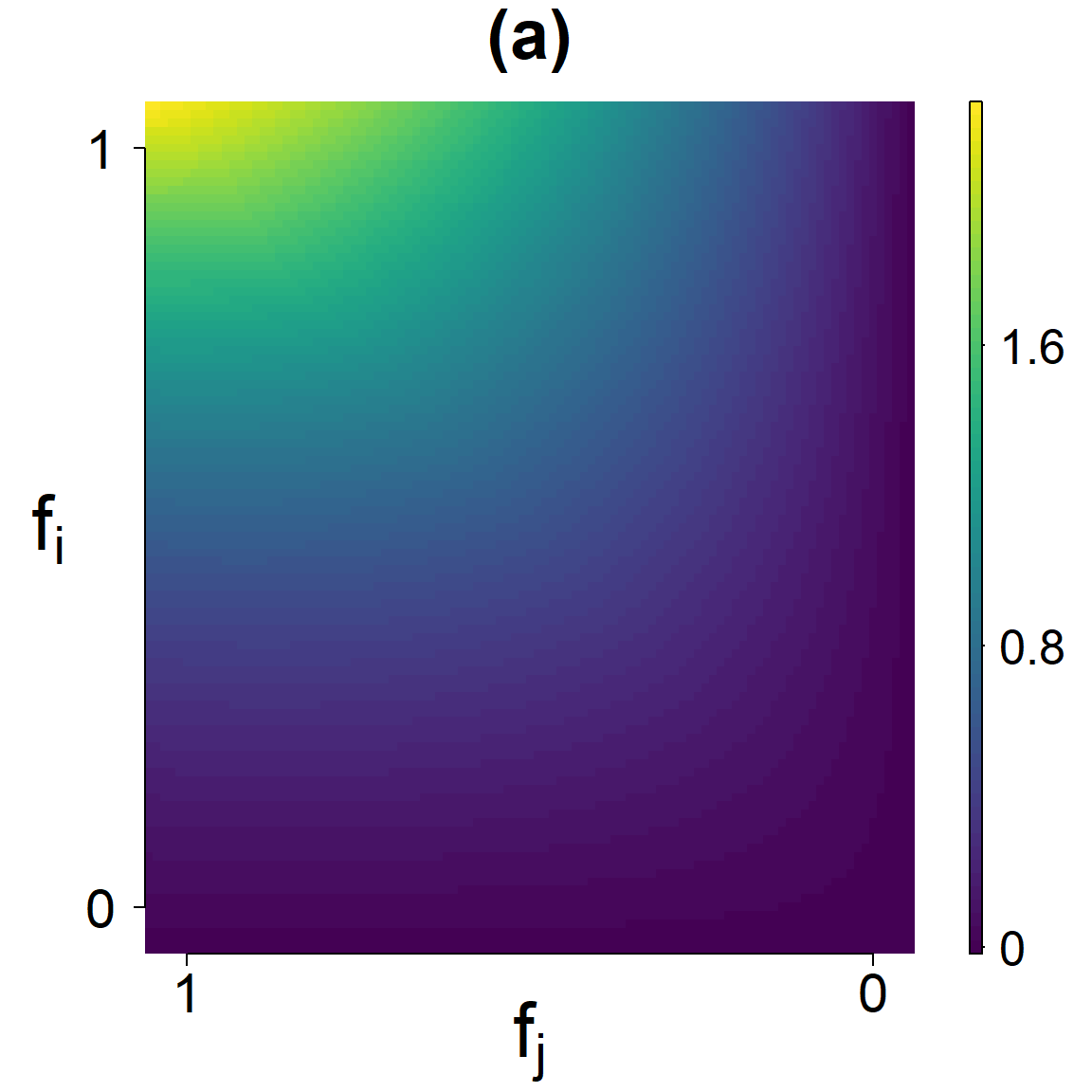}
 }
\subfigure
 {
 \includegraphics[width=0.46\linewidth,height=0.46\linewidth]{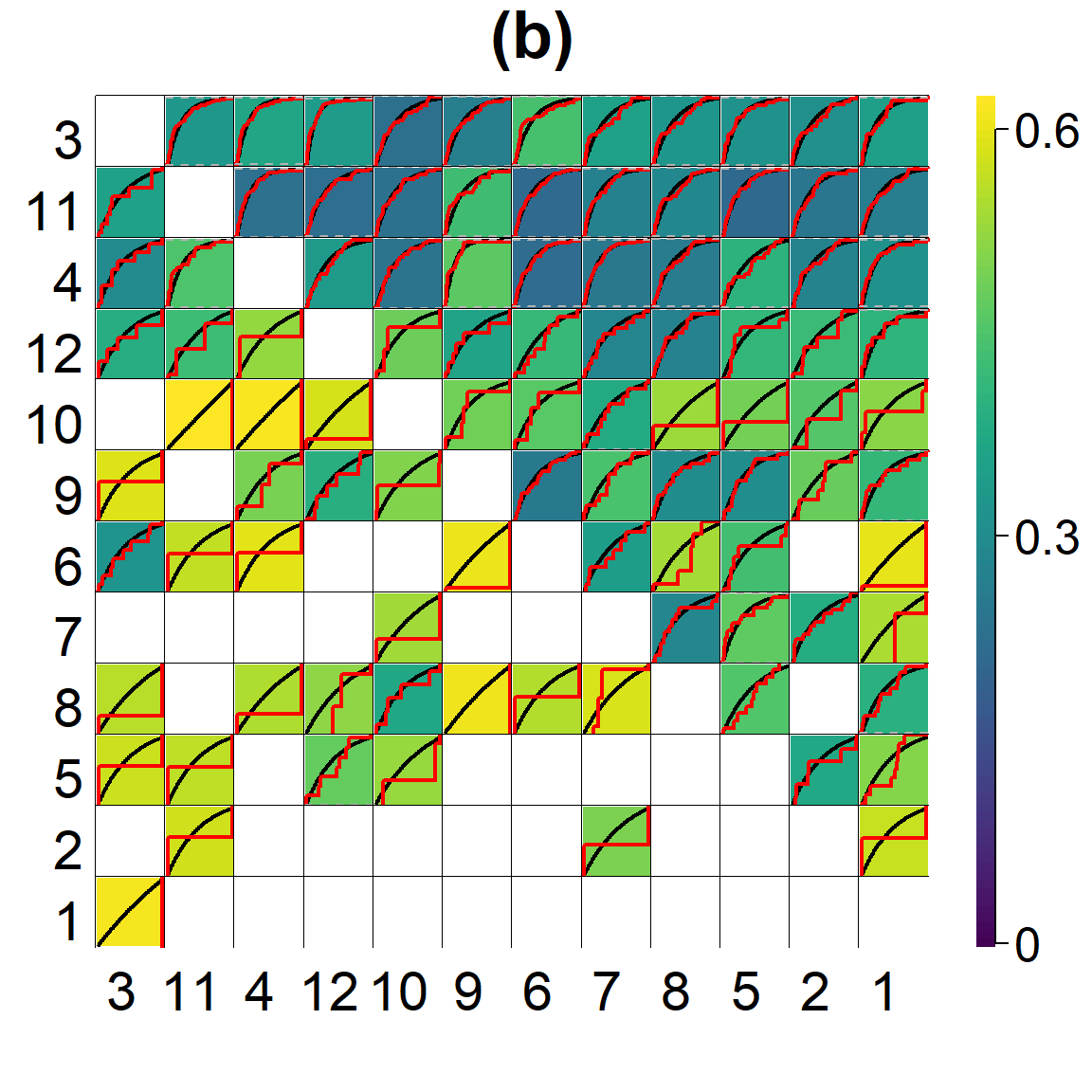}

 }
 \caption{(a) Contour plot for $\alpha^{i,j}:=g_\mathbf{\eta}(f_i,f_j)$ where 
 $f_i,f_j\in[0,1]$. (b) Matrix of K-S statistics after 
 fitting the C-HP model to the real data (reordered by I\&SI ranking). The rows and 
 columns of this matrix correspond to senders and 
 receivers of an agonistic behavior, respectively. Color shading reflects the values of 
 the K-S test statistics. Red lines are empirical 
 cumulative distribution functions of \textit{rescaled-inter-event}
 times and black lines are
 cumulative distribution functions of exponential random variable 
 with rate 1.}
 \label{fig_latent:model_1}
\end{figure}

Figure \ref{fig_latent:model_1}-(a) shows the 
contour plot of $g_\mathbf{\eta}(f_i,f_j)$, with $\eta_1=4.46,\eta_2=0.18$, and $\eta_3=1.46$, which are estimated values from real data analyzed in Section~\ref{sec_latent:real_data}. 
Note that the $x$-axis in Figure \ref{fig_latent:model_1}-(a) 
is decreasing from left to right, in order to be consistent
with the arrangement of a win/loss matrix, where the
interactions between the most dominant pairs are 
displayed in the top left. We notice that the
function takes higher values when $f_i>f_j$
(upper right triangle of 
Figure~\ref{fig_latent:model_1}-(a)), compared to values
when $f_i<f_j$ (lower left triangle of 
Figure~\ref{fig_latent:model_1}-(a)). This ensures that
winning interactions are directed more frequently
from a dominant individual towards a submissive individual,
mirroring the \textit{inconsistency} concept in I\&SI method.
The contour plot also shows that the form of this function
agrees with the idea of minimizing the total strength of
the inconsistencies in the I\&SI method: it has a smaller
value when $f_i<f_j$ and $|f_i-f_j|$ is larger
(moving from the diagonal to lower left 
triangle of Figure~\ref{fig_latent:model_1}-(a)).

Now, the intensity in the C-HP model is
$$\lambda^{i, j}(t)=
\lambda_1+g_\mathbf{\eta}(f_i,f_j)\sum_k\exp{(-\beta(t-t_k^{i,j}))}.$$

To assess the goodness-of-fit of point process models, according to 
the time rescaling theorem \citep{brown2002time}, we can test 
whether the \textit{rescaled-inter-event} times $\{ 
\Lambda_m:=\int_{t_{m-1}}^{t_{m}}\lambda(s)ds \}_{m=1}^{M}$, are 
independently distributed following an exponential distribution with 
rate 1. We fit this model to data corresponding
to interactions between a group of 12 mice,
using the inference procedure of Section~\ref{sec_latent:model_fitting}
We describe this data in more detail in Section~\ref{sec_latent:real_data}.
For each pair 
$(i,j)$, we conduct a Kolmogorov-Smirnov test on the 
\textit{rescaled-inter-event} times 
$\{ \Lambda^{i,j}_m:=\int_{t_{m-1}^{i,j}}^{t_{m}^{i,j}}\lambda^{i,j}(
s)ds
\}
_{m=1}^{M^{i,j}}$ and show the test statistics result in Figure 
\ref{fig_latent:model_1}-(b). The background color
indicates the value of 
the K-S statistics. 
This indicates good model fit for the C-HP model for 
the highly ranked animals (the
top three rows of (Figure~\ref{fig_latent:model_1}-(b)).
The values of these K-S statistics
increases, and there is evidence of a lack of fit
as we move from the top rows of 
Figure~\ref{fig_latent:model_1}-(b), particularly in the
lower left diagonal.
This is unsurprising as a majority of all interactions 
in this cohort are won by the 3rd animal (first row).
As such, in the C-HP model, these interactions lead to
a much larger value of $\lambda_1$ then would be
suited to describe interactions between other nodes.
This suggests that this model does not 
adequately address individual baseline event intensities when the
number of interactions between pairs can vary widely.
We introduce a correction to account for this next.

\subsubsection{Cohort degree-corrected Hawkes process (C-DCHP)}
\label{sec_latent:model_2}

\begin{figure}[ht]
 \subfigure
 {
 \includegraphics[width=0.33\linewidth]{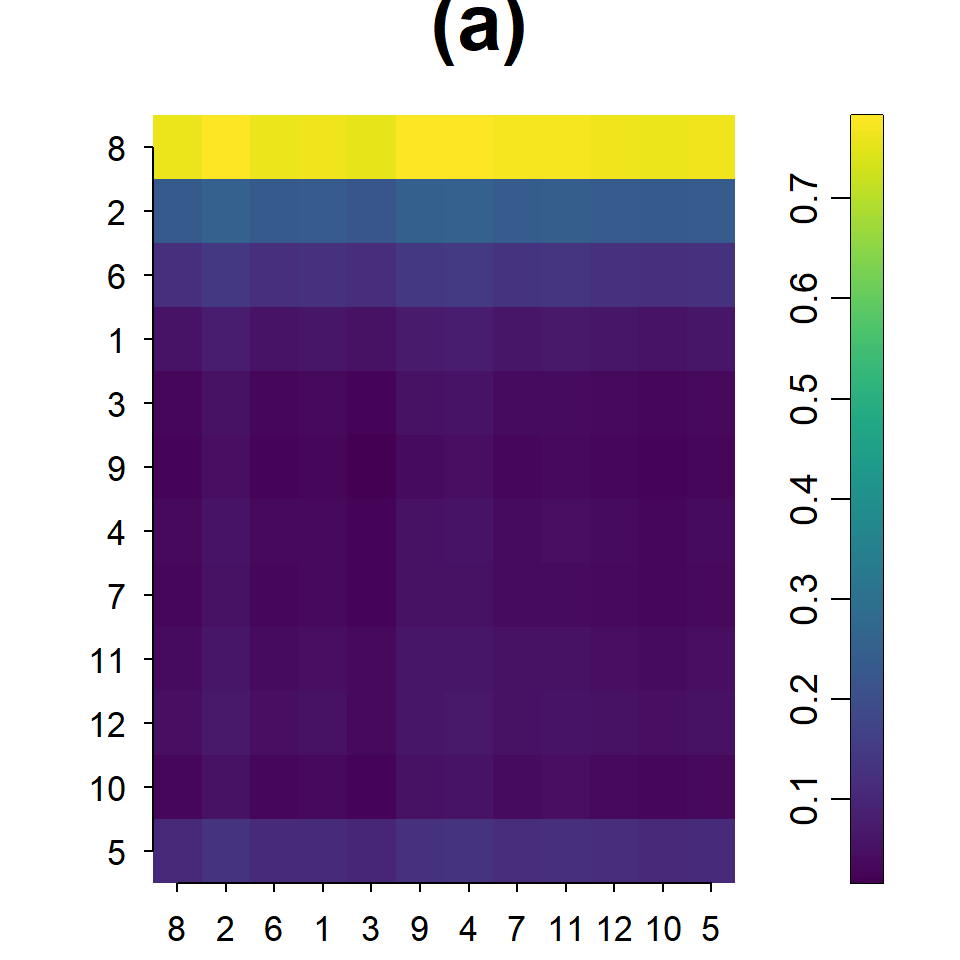}
 }
 \subfigure{
 \includegraphics[width=0.66\linewidth]{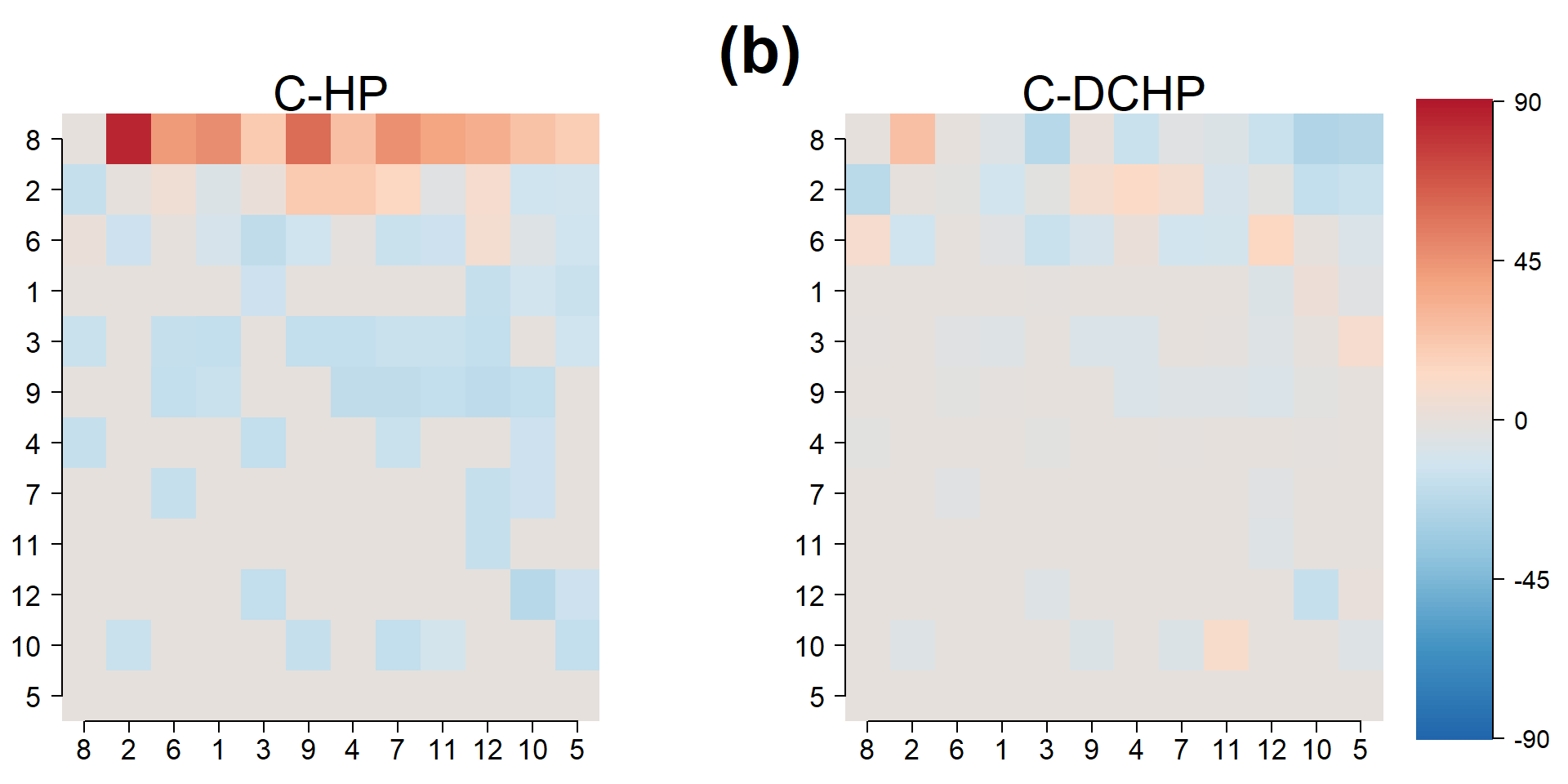}
 }
 \caption{(a) Matrix of baseline rates $\lambda_1^{i,j}$
 (reordered by I\& SI rankings). These degree-corrected
 baseline rates allow for a more flexible node level model,
 clearly seen in the top row (a mouse which is involved in starting a 
 large number of fights) and the bottom row (a mouse which 
 does not start any fights but is often fought).
 (b) Pearson residuals for the C-HP and C-DCHP models.
 }
 \label{fig_latent:model_2}
\end{figure}

The cohort Hawkes process model (C-HP) model assumes 
a constant
baseline rate $\lambda_1$ and is incapable of
modeling the degree heterogeneity of the observed nodes.
However, it can be observed from Figure \ref{fig_latent:model_1}-(b) 
that this model tends to consistently fit poorly for pairs which 
include certain individuals, for example those pairs in which the 
sender (winner) is individual 10 or the 
receiver (loser) is individual 11. To address 
this issue, we extend the C-HP model, allowing varying baseline intensity 
rates across pairs.
We accommodate degree heterogeneity in the pairwise baseline rate 
$\lambda_1^{i,j}$ by 
introducing a set of non-negative
out-degree-correction parameters
$\gamma_i$ and in-degree-correction parameters $\zeta_j$, 
$i,j=1,2,...,N$.
With the baseline rate defined as $\lambda_1^{i,j}=\gamma_i+\zeta_j$, 
we have the intensity function as 
$$
\lambda^{i,j}(t)=\gamma_i+\zeta_j+g_\eta(f_i,f_j)\sum_k\exp{(-\beta(t
-t_k^{i,j}))}.
$$ 
This model introduces degree-correction parameters in the baseline 
rate of the C-HP model, hence, we refer to it as the cohort degree-corrected Hawkes process model (C-DCHP). 
Figure \ref{fig_latent:model_2}(a) shows the baseline intensity 
matrix after fitting this 
model to the same cohort in 
Figure~\ref{fig_latent:model_1}-(b).
These estimates 
suggest that a model which allows more flexible intensities 
may indeed be needed to capture the heterogeneity in behavior across individuals.
We can see that this 
degree correction 
successfully adjusts for less active actors or recipients. For 
example, consider 
individuals $3$ and $6$; as shown in 
Figure~\ref{fig_latent:model_2}, their estimated 
$\lambda_1^{ij}$s are relatively large indicating that both 
individuals are more active in 
terms of interaction frequency; individual 6 tends not to win 
many fights and 
individual 3 tends not to lose many fights. Importantly, 
these differences in activity level are 
individual-level attributes and require separate 
consideration when we are interested in learning
about dominance hierarchy from dyad-level agonistic interaction data.
The resulting inferred ranks from the C-DCHP model for such
individuals are more 
consistent with the  
ranking obtained from existing methods,
compared 
to the inferred ranks from the C-HP model.
Similarly, to compare the fit of these models, we can look at the Pearson residuals
\citep{wu2021diagnostics}
from fitting each model to this data. We see, in Figure~\ref{fig_latent:model_2}(b)
that there is significant structure in the residuals after fitting the C-HP 
model, with large positive residuals for all interactions 
won by
the top
ranked animal. Fitting the C-DCHP model removes much of the
structure in these residuals and better captures the heterogeneous nature of
these pairwise interactions.
However, the
C-DCHP model consistently 
assigns a very large out-degree parameter value to the 
most dominant 
individual, as shown in Figure \ref{fig_latent:model_2}(a)
and seen for all real data examples we consider. 
This complicates inference for
the latent ranks of high-ranked individuals
and leads to poor model performance for any interactions not involving 
this individual. In observed data, 
these interactions
excluding this dominant individual
more often exhibit more sporadic behaviour, with 
long periods where no events are observed, a feature
we consider in the following model.


\subsubsection{Cohort Markov-modulated Hawkes Process (C-MMHP)}
\label{sec_latent:model_3}
In \cite{wu_markov-modulated_2021}, 
the Markov-modulated Hawkes Process (MMHP) 
is proposed to model 
general sequences of
sporadic and bursty event occurrences. The model 
utilises a latent two-state continuous-time Markov chain (CTMC) $Z(t)$
to better describe event dynamics.
In state 1 (the 
\textit{active} state), events occur according to a Hawkes process, 
while in state 0 (the 
\textit{inactive} state), according to a homogeneous Poisson 
process. The transition of $Z(t)$ is modeled through an infinitesimal generator 
matrix with parameters $\{q_1,q_0\}$, with the 
first row corresponding to transitions from the active state,
such that,
\begin{align}
\label{eq_latent:CTMC}
Q&=
\begin{bmatrix}
-q_{1} & q_{1} \\
q_{0} & -q_{0}
\end{bmatrix}.
\end{align}

Hence, for one MMHP, the conditional intensity function given the latent Markov process $Z(t)$, history events $\mathcal{H}(t)$ and parameter set $\Theta=\{\lambda_0,\lambda_1,\alpha,\beta,q_1,q_0\}$ is

$$
   \lambda(t|Z(t),\mathcal{H}(t),\Theta) = 
   \begin{cases}
        \lambda_0, & \text{when } Z(t)=0,\\
        \lambda_1+\alpha\sum_k\exp{(-\beta(t-t_k))}, & \text{when } Z(t)=1.
   \end{cases}
$$

Implicitly, the intensity function has the form
$$\lambda_0+(\lambda_1-\lambda_0)Z(t)+
\alpha Z(t) \sum_k\exp{(-\beta(t-t_k))}.$$ 
Thus, the latent process provides substantial flexibility
in modeling the baseline rate as well as the extent of historical event influence.

We can readily extend this model to the network setting, 
where for 
directed wins between
each pair $(i,j)$, the intensity follows,
\begin{align}
\label{eq_latent:network_intensity}
 \lambda(t|Z^{i,j}(t),\mathcal{H}^{i,j}(t),\Theta^{i,j}) = 
   \begin{cases}
        \lambda_0^{i,j}, & \text{when } Z^{i,j}(t)=0,\\
        \lambda_1^{i,j}+\alpha^{i,j}\sum_k\exp{(-\beta^{i,j}(t-t^{i,j}_k))}, & \text{when } Z^{i,j}(t)=1.
   \end{cases}
\end{align}
Here 
$\Theta^{i,j}:=\{\lambda_0^{i,j},\lambda_1^{i,j},\alpha^{i,j},\beta^{i,j}, 
q_1^{i,j}, q_0^{i,j}\}$ is the parameter set for pair $(i,j)$. 
$q_1^{i,j}$ and $q_0^{i,j}$ are the instantaneous transition probabilities for 
the latent CTMC $Z^{i,j}(t)$ of a pair $(i,j)$. $Z^{i,j}(t)$ are independent 
across pairs. The transition probability $q_1^{i,j}$ ($q_0^{i,j}$) represents the probability that pair 
$(i,j)$ transitions out of the active (inactive) state 
and is modelled as a function of the latent ranks,
$f_i, f_j$. To understand the behavior of these latent
state transition parameters for each pair $(i,j)$,
consider the stationary distribution of the latent CTMC, $Z^{(i,j)}(t)$. For an irreducible and 
recurrent CTMC $Z(t)$ with infinitesimal generator as shown in 
(\ref{eq_latent:CTMC}), a stationary distribution $\pi$ satisfies $\pi^TQ=0$ 
\citep{yin2012continuous}. Hence for a pair $(i,j)$, the limiting behavior of 
their latent state transitions dictates that they spend 
$\frac{q_0^{i,j}}{q_0^{i,j}+q_1^{i,j}}$ of their time in the active state, and 
all remaining time in the inactive state. With the hope that if $i$ dominates 
$j$, i.e. $f_i>f_j$, the pair $(i,j)$ will spend lots of time in the active 
state, we form the transition probabilities as, 
$$
q_1^{i,j} = \exp{(-\eta_3 f_i)}
$$ 
$$
q_0^{i,j} = \exp{(-\eta_3 f_j)}.
$$ 
Hence, when individual $i$ is stronger than individual $j$, 
node $i$ is more likely to start and continue winning
against node $j$
(i.e. large $q_0^{i,j}$ and small
$q_1^{i,j}$) 
than
to observe wins from $j$ to $i$.
This follows the
\textit{asymmetry} property of aggressive behavior 
in group animals. The limiting distribution of time
spent in state $1$ is $\mbox{Logistic}(\eta_3(f_i-f_j))$.

Given the latent process between pair $(i,j)$, $Z^{i,j}(t)$, 
we assume that $\beta^{i,j}$ is a 
constant $\beta$ across pairs.
In a similar vein to 
the C-HP and C-DCHP models,
we model the winner effect $\alpha^{i,j}$ 
as taking the form $\eta_1f_if_j \exp(-\eta_2 |f_i-f_j|)$.
The third component of the excitation parameter in
these first two models is now replaced
by the latent state indicator $Z^{i,j}(t)$.
As such, this equips the conditional intensity in the C-MMHP model
with additional flexibility,
alternating between a simpler homogeneous Poisson process
and a self exciting Hawkes process, with the limiting distribution of the process
aligning with the C-DCHP model.
Markov
modulation allows the self-exciting component of the intensity to dissipate
at points during the observation period,
a phenomenon that is often observed in animal behavioral studies where
interactions can be sparse at certain times.
In other words, this modulation is introduced to account for the
often seen sporadic nature of interactions between animals,
arising from initial exploratory aggressive interactions and 
later interactions which may determine the precise hierarchy,
before potentially stabilising.

Like the C-DCHP, we again consider
the degree correction described in the previous model here,
giving $\lambda_{0}^{i,j} = \gamma_i + \zeta_j$.
$\lambda_{1}^{i,j}$ is defined by
$$
\lambda_{1}^{i,j} = \lambda_{0}^{i,j}(1 + w_{\lambda}),
$$
for a common $w_{\lambda}\geq 0$, to ensure that
the base rate of the point process in the
active state is greater than the inactive state.
Hence we have the intensity from $i$ to $j$ given by
$$\lambda^{i,j}(t)=
\lambda_0^{i,j}+
(\lambda_1^{i,j}-\lambda_0^{i,j})Z^{i,j}(t) + \eta_1f_if_j
\exp(-\eta_2 |f_i-f_j|)
Z^{i,j}(t)\sum_k\exp{(-\beta(t-t_k^{i,j}))}.$$

{
This proposed model therefore provides a potential mechanism to describe the 
establishment of a realised dominance structure,
as an expression of a latent hierarchy, by incorporating each of the components of 
the C-DCHP model
and allowing Markov modulation. This is a key step towards better understanding the
true process behind such interactions, a key question highlighted
by \cite{williamson2016temporal}.
The mechanism we propose here agrees with existing methods from the animal 
behavior literature and captures properties commonly seen in data of this 
form. For example, bursty behavior is commonly seen in aggressive and 
subordinate interactions \citep{lee2019temporal},
which is readily incorporated into our model through the use of a Hawkes 
process. Through the parameterization of the Hawkes process, our proposed 
model is also designed to capture a linear hierarchy. This agrees
with many of the existing methods in this area. However, these existing
aggregate methods fail to adequately account for wins
in unexpected directions, such as from lower ranked to higher ranked
animals, which can and do occur.
Interactions such as these are unsurprising, particularly
when the animals are first placed together, as they fight
to gain social information about the social hierarchy. These
interactions may continue to occur also.
It can be advantageous for animals 
to improve their position by beating similarly ranked animals,
to better obtain limited resources and potentially dissuade
lower ranked animals from being aggressive towards 
them.
While this is only seen among strong mice, in other species it occurs
throughout the hierarchy \citep{hobson2021aggression}.

We observe this 
only among the top ranked mice in our data,
where the top ranked animal
can be defeated late in the 
observation period \citep{williamson2016temporal}.
By equipping each directed interaction pair with a point process,
there is always some likelihood that interactions will be observed between 
animals in an uncommon direction. Our C-MMHP model goes further than this. 
With a Markov Modulated Hawkes Process, we allow the likelihood for 
interactions to vary in time, alternating between a Poisson process and a 
Hawkes process. This additional flexibility is well suited for capturing 
interactions in an uncommon direction. In particular, interactions of this 
type are often sporadic, often occurring after long periods where there are 
no interactions between that pair or in that direction.

\subsection{Model inference}

\paragraph{Bayesian modeling.} Throughout this paper,
we adopt a 
Bayesian framework for our model inference. Assuming a prior 
distribution for the model parameters and given a model 
likelihood, the posterior distribution for quantities of interest 
can help us calibrate the uncertainty in the model. This is an 
important aspect of our current research strategy. First, we need 
tools that can quantify uncertainty in rank inference. For 
example, in \cite{williamson2016temporal}, the analysis shows that
the \textit{pair-flips} phenomenon exists in some cohorts, which 
means that the direction of aggressive interaction changed over 
time. In a Bayesian modeling framework, we can naturally capture 
this effect through uncertainty in the model parameters: we 
suspect that individuals that are involved in pair-flip phenomena 
should have larger posterior variances for their latent ranks. 
Second, in each cohort, there always exists some pairs that have 
few or no interactions across the observation time window.
A Bayesian framework can achieve robust inference in such 
conditions, with the assistance of prior assumptions and by 
borrowing strength from the data of other pairs. In this paper, we
will use the Stan modeling language \citep{carpenter2017stan} to 
fit all models and to obtain posteriors samples for model 
parameters. We describe further
details of our inference procedure
in Section~\ref{sec_latent:model_fitting}.

\section{Results}
\label{sec_latent:results}

\subsection{Comparison Models}
Before analysing the performance of our
models on both real and simulated data, we first
describe in detail existing methods used to
analyse dominance behavior in animals, which we shall use for comparison 
of inferred rankings.
As described above, these methods can be broadly 
classified as \textit{functional} and \textit{structural}.

\subsubsection{Functional methods}
\label{sec_latent:related_functional}
Along with the I\&SI method,
\cite{so2015social} and \cite{williamson2016temporal} use the 
Glicko rating system to 
calculate temporal changes in dominance scores of each animal 
in each cohort. This 
is a dynamic paired comparison system that calculates a 
temporal 
sequence of cardinal scores based on the history of dyadic 
wins and losses 
\citep{glickman1999parameter}.
All individuals start with the same initial rating. 
After each observed fight between a pair, the winner (or the 
loser) gains (or loses) 
points according to a decreasing function of the difference 
between their previous 
scores. In this case, fighting between pairs whose scores 
differ a lot will not result 
in significant changes in the system. The calculation of the 
Glicko score depends 
on a predefined constant, which determines the volatility of 
the score changes. Since 
scores are computed after each fighting event, 
this method can
capture the 
temporal dynamics of the dominance hierarchy,
although it does 
not
account for temporal components, such as the time between 
events.
\cite{williamson2016temporal} also 
provides a clear visualization of the change in the dominance 
score based on this 
method, where the emergence and stabilization of the
hierarchy 
can be easily deduced from 
the graph. However, this method is ad-hoc in the sense that 
there are no theoretical 
rules for researchers to choose important key aspects of this 
method, including the initial rating, the decreasing function 
for 
changing a pair's scores after an observed fight, or the 
constant controlling the 
volatility of score changes. Since the method focuses on 
summarizing the observations 
without any formal modeling, it can be hard to provide formal 
insights regarding the 
evolution of hierarchy dynamics. 
It is also not always clear 
how to draw a conclusion 
about the hierarchy structure
from the visualization of the rating system. 

\subsubsection{Structural methods}
\label{sec_latent:related_structural}
\cite{lindquist2009data} apply winner-loser 
models to real experimental data of hens and show the lack of fit between these 
models and the data. However, this procedure is qualitative only,
by comparing simulation 
results from the models with the real data. The 
probabilistic generative 
models we proposed in Section~\ref{sec_latent:models}
are able to capture these important animal behavior 
phenomena, including the \textit{winner effect, bursting and pair-flips}. 
We also
develop a corresponding statistical inference procedure which means that 
model-fitting can be assessed by rigorous statistical model diagnostics, rather 
than relying on simulations as in \cite{lindquist2009data}. We analyse our
models using these diagnostics in Section~\ref{sec_latent:sim_results}
and Section~\ref{sec_latent:real_data}.

A more recent structural model is
\cite{de2018physical}, which
introduced a physics-inspired model to infer cardinal 
hierarchical rankings of individuals in directed networks. By assuming that 
individuals are more likely to interact with others of similar rank, they propose 
an optimization solution and a generative model to find real-valued ranks of 
individuals. 
For a pair $(i,j)$, with latent rank variables
$f_i$ and $f_j$, the aggregate-ranking model uses Poisson regression to model 
the aggregate counts between the pair over the entire observation period, 
denoted as $N_{i,j}$, as a function of the difference in their ranks.
This is essentially the counting process evaluated at time $T$, $N^{i,j}(T)$, 
ignoring the exact event times. 
The only information used in the model is the existence and 
direction of the interactions in the network. We refer to this model as the 
\textit{aggregate-ranking} model. Only using the aggregate counts of interactions 
makes it hard to address phenomena like the \textit{winner effect, bursting and 
pair-flips} mentioned in \cite{lindquist2009data}. Event time data which records 
when the aggressive behaviors occur is highly detailed and contains
information needed to describe the important phenomena mentioned earlier. 


\subsection{Model implementation}
\label{sec_latent:model_fitting}

To perform inference for each of these models we perform Bayesian inference
using the Stan programming language \citep{rstan_20}.
We impose weakly informative priors for the
model parameters where possible.
In particular, for each
model we use $\mbox{half-N}(0,1)$
priors on $\eta_1,\eta_2,\eta_3$ and $\beta$, $\mbox{U}[0,1]$
priors for $f$, and a $\mbox{half-N}(0,1)$ 
prior for 
$w_{\lambda}$ in the C-MMHP model,
to ensure that the rate in the active state
is greater than in the inactive state.
For the degree-corrected
models we place Laplace priors
on $\gamma$ and $\zeta$, to account for nodes which do not win 
or lose any fights.
Full details of the inference procedure to infer
the latent states of the C-MMHP model are given
in \cite{wu_markov-modulated_2021}.

\subsection{Synthetic results}
\label{sec_latent:sim_results}
We first wish to validate our proposed models using
simulated data where we aim to recover the
{\em ground truth} latent ranking vector.
To compare the three proposed models,
we simulate 50 independent C-MMHPs
generating event winning times between
10 nodes with uniformly separated
true latent ranks.
By fitting the synthetic data 
with
our proposed three models,
we can obtain the inferred latent ranks as shown in 
Figure~\ref{fig_latent:simulation}-(a). Inference using C-MMHP 
and the C-DCHP
both 
recover the true latent ranks well, with the
C-HP showing considerable bias for several nodes. It is unsurprising that
there is little difference between the
inferred rankings from the C-DCHP and C-MMHP model,
given their similar limiting distribution.
Figure~\ref{fig_latent:simulation}-(b) shows 
an example of estimated intensity for one pair of individuals 
in one simulated process (from the top ranked to second ranked individual), 
indicating that the C-HP and C-DCHP models cannot
capture the true intensity as well as the C-MMHP
model. The C-HP incorrectly captures the true decay parameter
while both the C-HP and C-DCHP 
model overestimate the intensity repeatedly for events
which occurred in the inactive state, with both showing more over and underestimation, althought the C-MMHP does
incorrectly classify some events.

We also compare the inferred ranking from each of
these models and 
the comparison 
models discussed previously with the known true ranking. We summarize
this 
using the Spearman rank correlation between the estimated 
ranking obtained from each method and the true ranking, as shown in
Figure~\ref{fig_latent:simulation_residual}. We note that
the C-MMHP model recovers the true ranking
best, with the C-HP and C-DCHP models
also performing well but showing more variability. 
As a structural method,
the I\&SI model recovers the true ranking reasonably well for simulations from the generative C-MMHP model.
In this scenario a large proportion of fights occur
in the active state of the simulated C-MMHP model
and so agreement between the I\&SI method and our 
proposed models is expected. 

\begin{figure}[h!]
 \subfigure
 { 
 \centering
 \includegraphics[width = 0.65\linewidth]{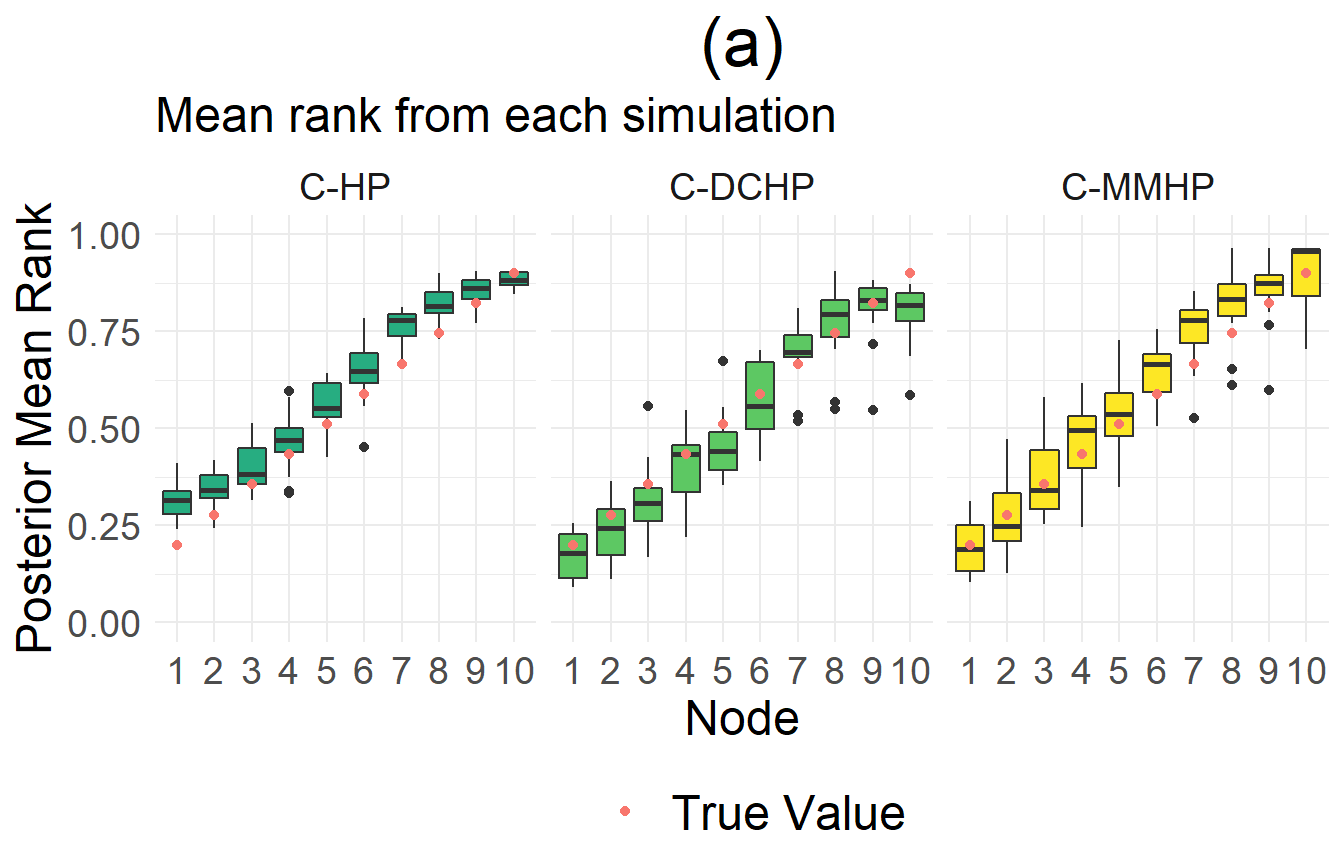}
 }
 \subfigure
 {
 \centering
 \includegraphics[width =0.65\linewidth]{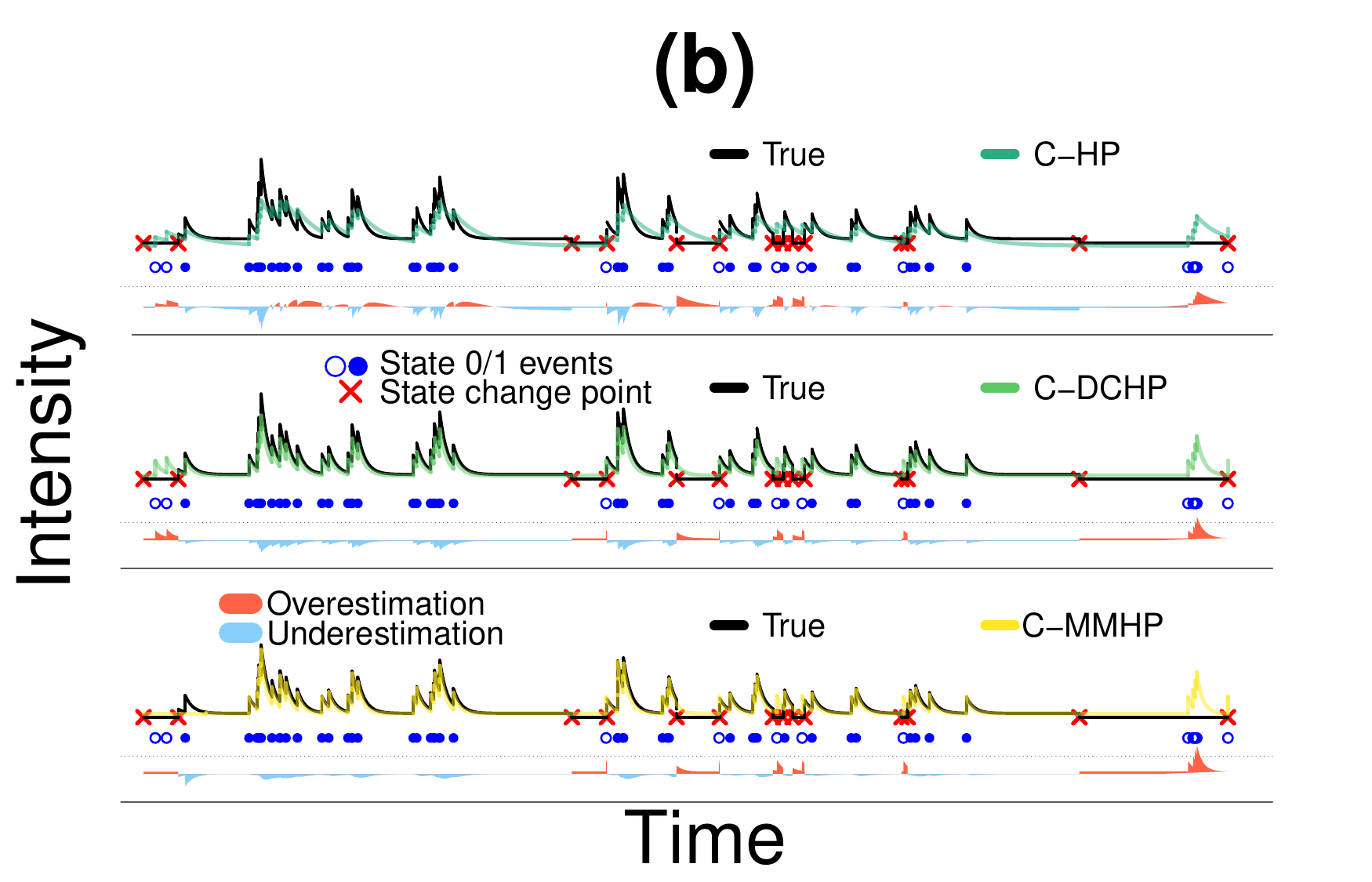}
 }
 \caption{Simulation results from
 simulating 50 interaction datsets with common C-MMHP parameters.
 (a) Shows posterior inference 
 of latent rank variable $f_i, i=1, \dots ,10$ by C-HP, C-DCHP and C-MMHP. 
 Each value is the posterior mean for $f_i$ inferred from 50 independent
 simulations from a C-MMHP model with the same underlying parameters, with the
 true rank values overlaid in red.
 (b) Show the inferred intensity for one pair of 
 individuals (top ranked to second highest ranked)
 in one simulation using three models. 
 Here we fit each of the C-HP, C-DCHP and C-MMHP models to this
 data and plot the inferred intensity function for this pair.
 The events and the state they occurred in, along
 with the times the process changed state, are
 also shown in this plot.
 The
 red/blue shaded area underneath shows the
 magnitude of the error in the estimation of the intensity in terms of overestimation
 and underestimation.}
 \label{fig_latent:simulation}
\end{figure}

\begin{figure}[ht]
\centering
 \includegraphics[scale = 0.5]{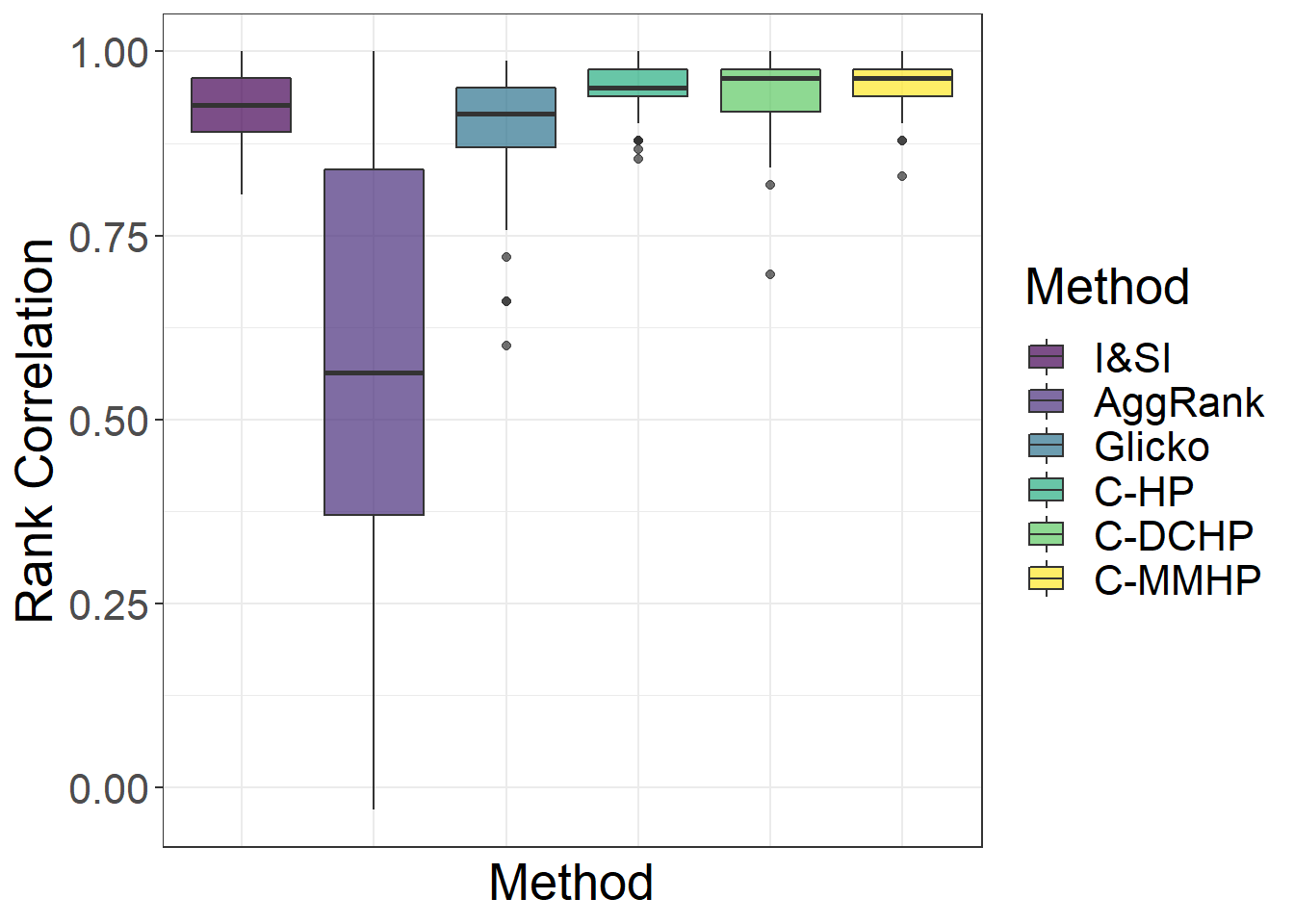}
 \caption{Shows the Spearman rank correlation between the
 inferred ranking from each of these models, along with existing
 methods, and the known
 true ranking.}
 \label{fig_latent:simulation_residual}
\end{figure}

\subsection{Real data results}
\label{sec_latent:real_data}

We next 
fit our models, C-HP, C-DCHP and
C-MMHP to the ten mice cohorts studied by 
\cite{williamson2016temporal}, which
consisted of placing each cohort of twelve male mice
in a large custom built vivarium. Intensive 
behavioral observations were conducted for one
to three hours per day during the dark cycle
over twenty-one consecutive days. Trained
observers recorded all occurrences of the
behaviors (including fighting, chasing,
mounting, subordinate posture and induced-flee).
The details of each behavioral event are also
recorded, 
identifying the actor which wins the interaction and
the actor which loses the interaction, along with the
timestamp and location. 

Using various measurements in social hierarchy
analysis and social network 
analysis, \cite{williamson2016temporal} demonstrates
that these mice cohorts form significantly linear social
dominance hierarchies. The work also examines the
temporal changes in the mice social hierarchy and
shows that in most of the ten cohorts, the dominance
hierarchies emerge rapidly and become stable by
the end of the second week. Although results of
the quantitative analysis are thoroughly discussed and the
patterns in the temporal dynamics are summarized
qualitatively, there are still observations in some
cohorts that disagree with the authors' speculations,
which have been unexplained in existing work but are naturally
accounted for in our model.
We have addressed how our model provides a potential
mechanism for
the connection between the establishment of a latent ranking 
of these animals with the observed social interactions
and described how
it captures interactions in an uncommon direction.
We will verify that our
model does describe this data well using multiple metrics,
illustrating
that a flexible probabilistic 
model incorporating these
hypothesised processes can be constructed. We then also show how
further information may be available from these models, such as
information about the distribution of individuals' dominance power.
We also compare the results from fitting
the following existing models to this data: a dynamic
social network in latent space model,
and a Markov-modulated Hawkes process without incorporating
any latent ranking structure between the nodes.
We first briefly describe these models.
Additional analysis results for each individual cohort
are available in a supplemental file.
\paragraph{Dynamic social network in latent space model 
(DSNL)\citep{sarkar2006dynamic}.} This model is constructed for 
dynamic network data with binary links which is observed in
discrete time steps. The model associates each node in the
network with a latent space variable that can move in discrete time,
and specifies that the move is Markovian. For node $i$ at discrete time $d$,
the latent variable is denoted as $f^{(d)}_i$. We tailor this
model to our observed mice interaction data by changing the 
binary link assumption in the original model to allow for 
aggregate counts by using a Poisson link instead of a 
logistic link. We construct discrete time steps to be 
the ending time of each day in the observation time window,
i.e. $t^{(d)}$. Hence, for each pair $(i,j)$, we have the
count of their interactions during day $d$, $N_d^{i,j}:=N^{i,j}(t^{(d)})-N^{i,j}(t^{(d-1)})$, where
$N^{i,j}(t)$ is the counting process for pair
$(i,j)$ evaluated at time $t$. Further details of this model will be omitted here. 

\paragraph{Markov-modulated Hawkes process without network ranking structure
(I-MMHP) \citep{wu_markov-modulated_2021}.} In this model, we assume that 
the intensity function of (\ref{eq_latent:network_intensity}) allows for 
different parameter values $\Theta^{i,j}$ across pairs.
This means there is no structure between nodes and a latent ranking of the
animals cannot be inferred.
The independent 
structure of the parameters in this model 
is less constrained than
our C-MMHP model, where we consider network structure 
between nodes
to learn latent rankings.

\paragraph{Summary measures for evaluating model performance.}
Our real data analysis results will be summarized from several perspectives:
(i) inference for the latent ranks 
and a summary 
of the properties captured by each of the different methods, 
prediction performance, both in terms of  (ii) predicted
events and (iii) predicted evolution of dynamics over time and
(iv) additional insights available through the C-MMHP model,
which may provide potential future research directions.
We compare the results of the C-HP, C-DCHP and C-MMHP models
under the first three of these perspectives. Because
the nature of the three other comparison models - aggregate-ranking,
DSNL and 
I-MMHP - differs, they are fitted and compared from different perspectives. 
The aggregate-ranking model (and also the I\&SI 
method we discussed previously) estimates a 
static ranking and will be discussed in terms 
of inference for the latent ranks only. The 
I-MMHP is a point process model and can 
be evaluated using the same point process methods as our 
latent ranking point process models. However, the I-MMHP cannot be used
to infer a latent ranking. Finally,
both the DSNL and 
I-MMHP models can perform prediction of events (or event counts)
and can serve as comparison models in the 
prediction performance section.

\subsubsection{Inference on latent rank}

We fit the 
C-HP, C-DCHP, C-MMHP and aggregate-ranking models
to our data of ten cohorts separately. 
Figure~\ref{fig_latent:real_rank} shows the relationship
between I\&SI rank and posterior 
draws of latent ranks using our three models and
aggregate-ranking model in two cohorts,
Cohort 5 in Figure~\ref{fig_latent:real_rank}~(a)
and Cohort 3 in Figure~\ref{fig_latent:real_rank}~(b).
While we have ordered the animals by their estimated I\&SI 
ranking, this is not to take it as some reference ranking
but to highlight how our models capture a different
ranking structure.
These cohorts display the two types of common characteristics we observe
across the 10 cohorts. 
Cohort 5 provides an example of general behavior seen in 7 of the 10 cohorts 
studied here. In this example, there is general agreement between the rankings
inferred from each of the Hawkes models, with the C-MMHP model displaying less
uncertainty than the alternative models.
The C-MMHP model identifies 3 approximate groups within the rankings, which
is seen in several cohorts.
As such, the bursty model assumption seems to agree with the dynamics
evident here, with these bursty dynamics largely aligning with the power hierarchy. 
We also see animals such as \# 7 and \# 10 where the ranking from the C-MMHP model deviates
from the I\&SI ranking, as these animals are 
involved in relatively few fights and as
such there are few periods of bursty fights. 
In this cohort all animals lose a similar
number of fights, with the percentage of all fights which are 
classified as active by the Markov modulation being
very high for the
whole cohort.

Different from the seven cohorts represented by cohort 5, Cohort 3 and another two
cohorts exhibit another form of common behavior.
Here it is difficult to identify differences in the rankings for
many of the animals. There is also
weaker agreement between the C-MMHP model 
and the C-HP/C-DCHP models, while we also see disagreement between
the I\&SI ranking and the simplest of our models, the C-HP model.
Interestingly, here all inactive originate from animal \# 5, which
is ranked significantly lower by the C-MMHP model than its I\&SI ranking.
Similarly, this animal has a large relative out degree estimate
under both the C-DCHP and C-MMHP models. For these cohorts there
are a large number of sporadic fights and a large percentage of 
wins attributed to
a single animal, who seems to indiscriminately 
win fights against
all other animals repeatedly. Although this behavior is
rewarded by traditional ranking methods, it is not clear that
this should be an indication of dominance, with other animals
perhaps unable to learn social information about the losers of these
fights \citep{hobson2020differences}. Our C-MMHP model
places less emphasis on this behavior, giving such animals
a lower ranking than existing methods.

Given that the Aggregate Ranking method and the 
I\&SI method use the same win/loss matrix information,
it is expected that they show strong agreement.

\begin{figure}[ht]
 \subfigure
 {
\includegraphics[width=0.48\linewidth]{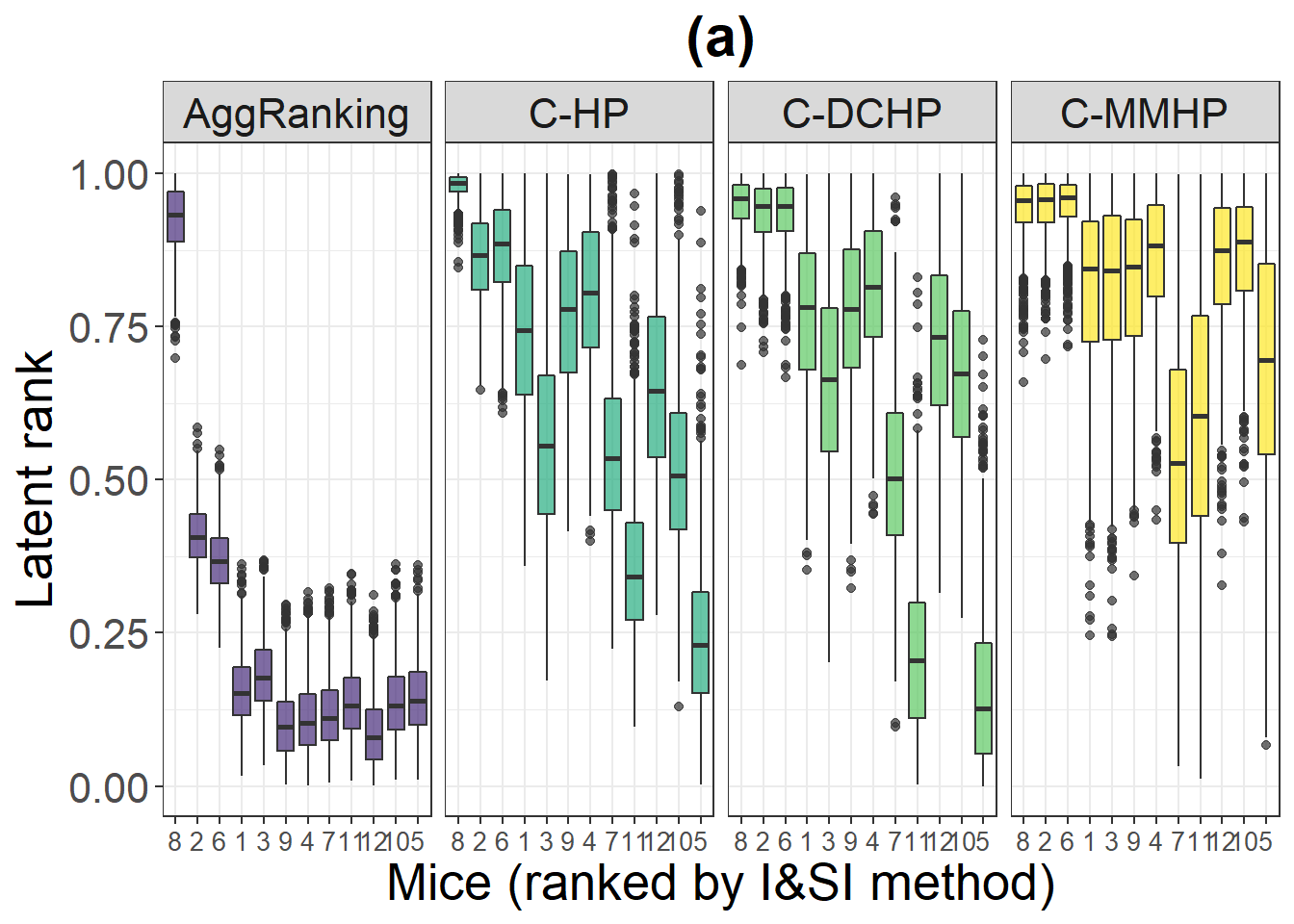}
 }
 \subfigure
 {
\includegraphics[width=0.48\linewidth]{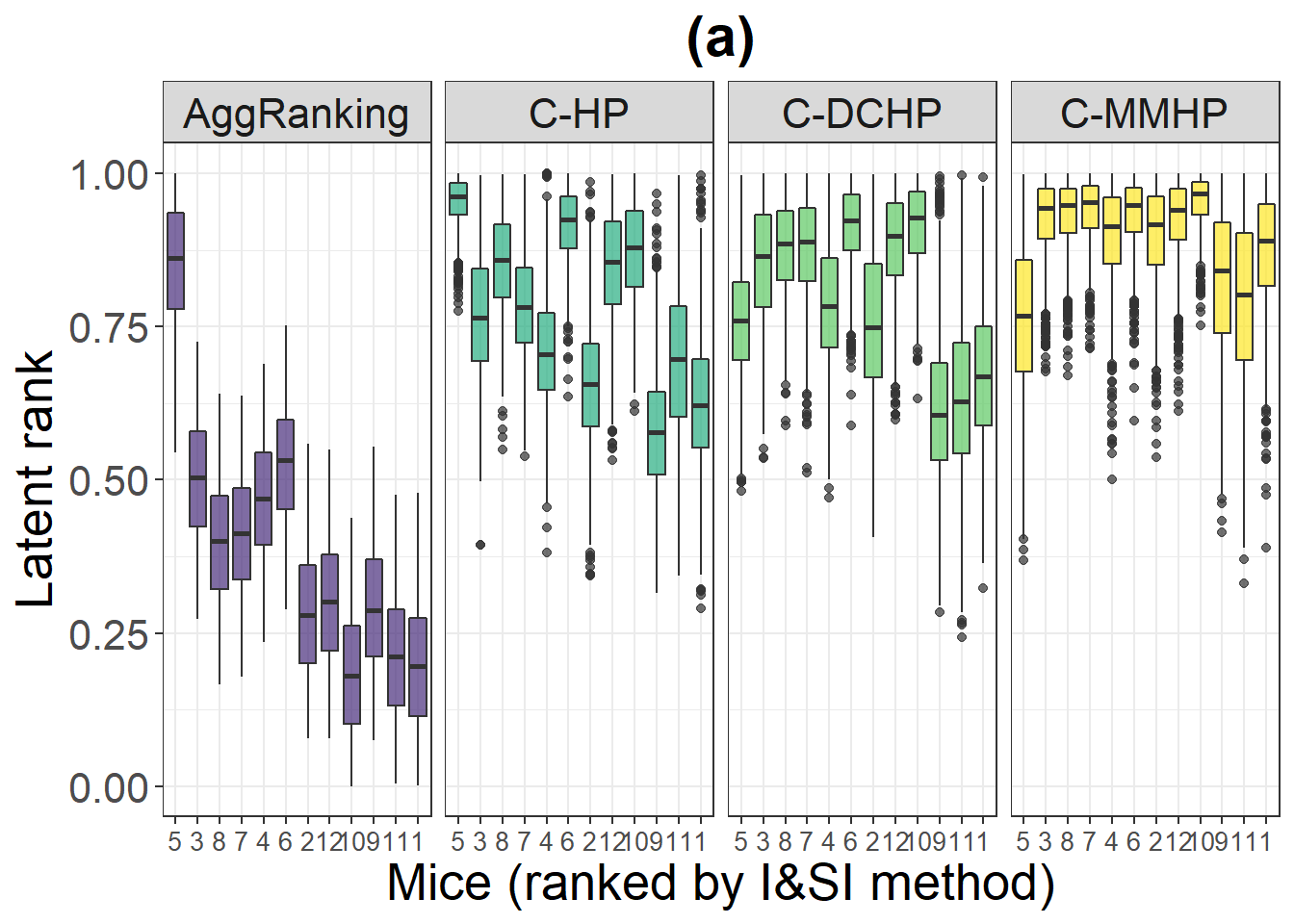}
 }
 \caption{Real data fitting results. (a) Comparison of rank inference using different model with I\&SI rank for Cohort 5. (b) Comparison of rank inference using different model with I\&SI rank for Cohort 3.}
 \label{fig_latent:real_rank}
\end{figure}

\subsubsection{Prediction}

We can also use posterior predictive distributions to validate the 
models considered.
For each model, we split the data into two time periods: (1) the first 15 days 
of data, $\mathcal{H}^{i,j}(t^{(15)})$, where $t^{(d)}$ is the ending 
observation time for the $d$-th day, which is used to estimate the model and 
(2) a prediction window from day 15 to day $t^{(d)}$, for $d=16,...,21$,
the remaining observation period,
which 
allows us to compare models across different prediction horizons. For each 
prediction horizon $t^{(d)}$, we generate a predicted point process separately 
over the time period $t^{(15)}$ to $t^{(d)}$, given each posterior draw of 
parameters and the historical events in the first 15 days. 
Hence, the predicted counting process $\hat{N}^{i,j}(t)$ is constructed by generating processes in 
each prediction period and adding these to the true process in the 
model-fitting period.
For each prediction horizon and model, we generate 1000 
posterior processes, corresponding to 1000 posterior draws from the 
posterior 
distribution for the model parameters. Following \cite{sarkar2006dynamic},
we 
can also make predictions over these same time windows using the DSNL 
model.

Two aspects of the predictions are evaluated, the accuracy of predictions 
for 
the interaction counts and the prediction of the rankings.

For each point process model and for each different prediction horizon 
$d=16,...,21$, the number of total interactions for pair $(i,j)$ during 
the 
prediction period can be estimated by 
$\bar{\hat{N}}^{(i,j)}(t^{(d)})-N^{(i,j)}(t^{(15)})$, where 
$\bar{\hat{N}}^{(i,j)}(t^{(d)})$ is the average count of 
wins
across 
1000 posterior processes. We arrange the prediction counts in a matrix 
$\hat{A}^{(d)}$ such that each $(i,j)$ entry is the predicted number of 
interactions for pair $(i,j)$ from the end of the 15th day until the $d$th
day.
To quantify the accuracy of these predicted counts, we use the
mean absolute error (MAE)
of the difference between the estimated and real win/loss matrix 
$A^{(d)}$, 
$$\frac{1}{n}\sum_{i,j}| \hat{A}^{(d)}_{ij} - A^{(d)}_{ij}|
$$
The smaller the MAE, the
closer the model's predictions of the interaction counts are to the 
observed 
data. Figure 
\ref{fig_latent:real_predict_rank}-(a) summarizes the result 
across 
all cohorts,
by taking the median predicted counts for each pair across each of 1000
posterior draws.
The C-MMHP, C-DCHP and I-MMHP
models provide the best predictions
of interaction counts, with the smallest MAE across all 
prediction horizons, with C-MMHP slightly outperforming the other models.

We also infer a proxy predicted rank of individual $i$ at prediction time $t^{(d)}$ by 
introducing the out-degree intensity  
$$
\hat{\lambda}_i(t^{(d)})=\sum_j\hat{\lambda}^{i,j}(t^{(d)}).
$$ 
The Glicko 
score ranking system serves as a bench mark for us to compare to, 
as it is a dynamic score.
While this is not a true dynamic score, we can construct it for each
of the models we consider here, allowing us to use it as a comparison across them, identifying agreement between each of these models and the
evolution of the Glicko ranking.
We compute the Spearman rank 
correlation of our 
inferred rank with the Glicko score at the end of the prediction day. 
Figure 
\ref{fig_latent:real_predict_rank}-(b) summarizes the result for all 
cohorts. 
The C-MMHP model closely aligns with the Glciko
score, with rank correlation close to 1,
while the unconstrained
I-MMHP model also performs well in this
scenario.

\begin{figure}[ht]
 \subfigure
 {
 \includegraphics[width=0.47\linewidth]{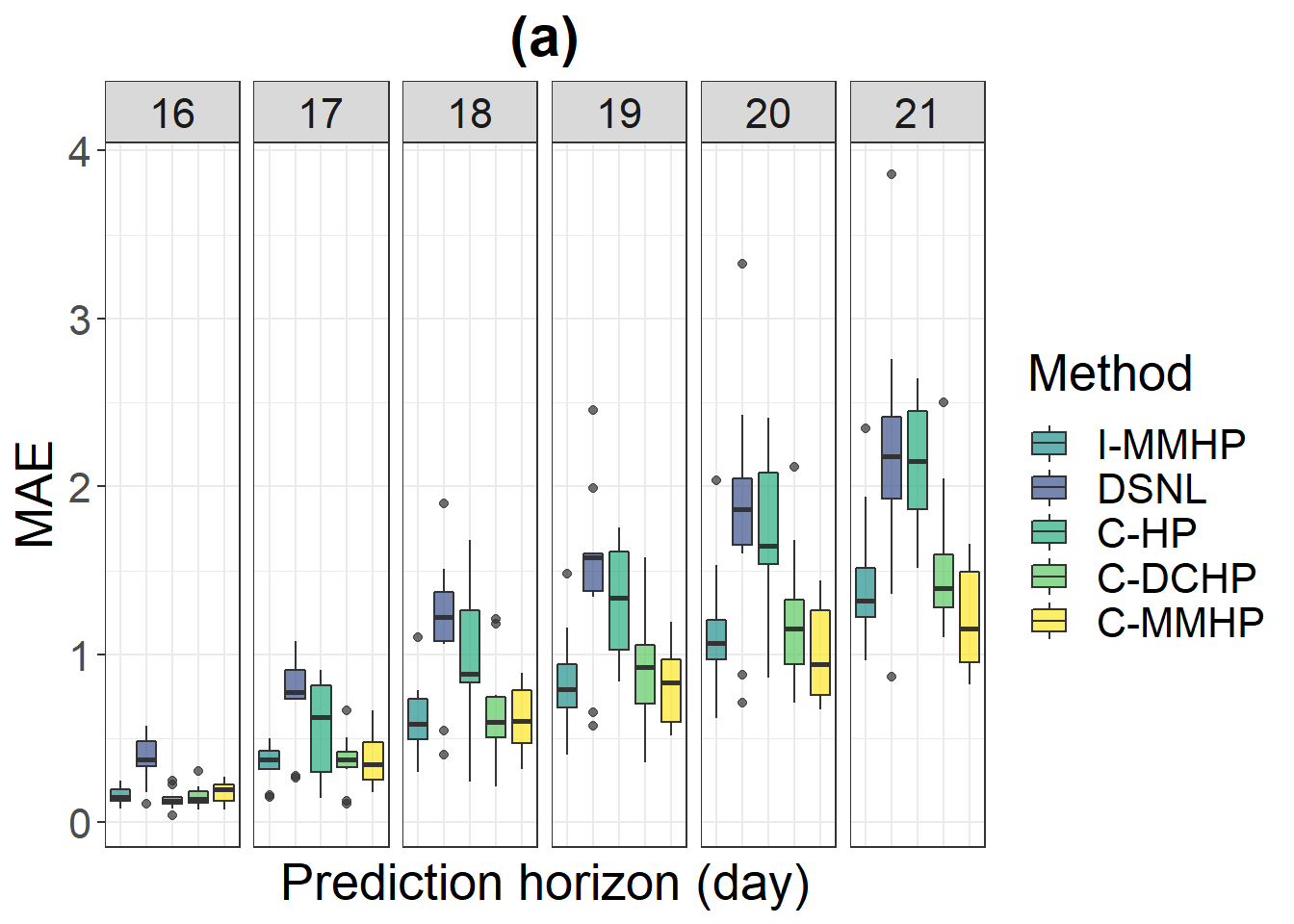}
 }
 \subfigure
 {
\includegraphics[width=0.47\linewidth]{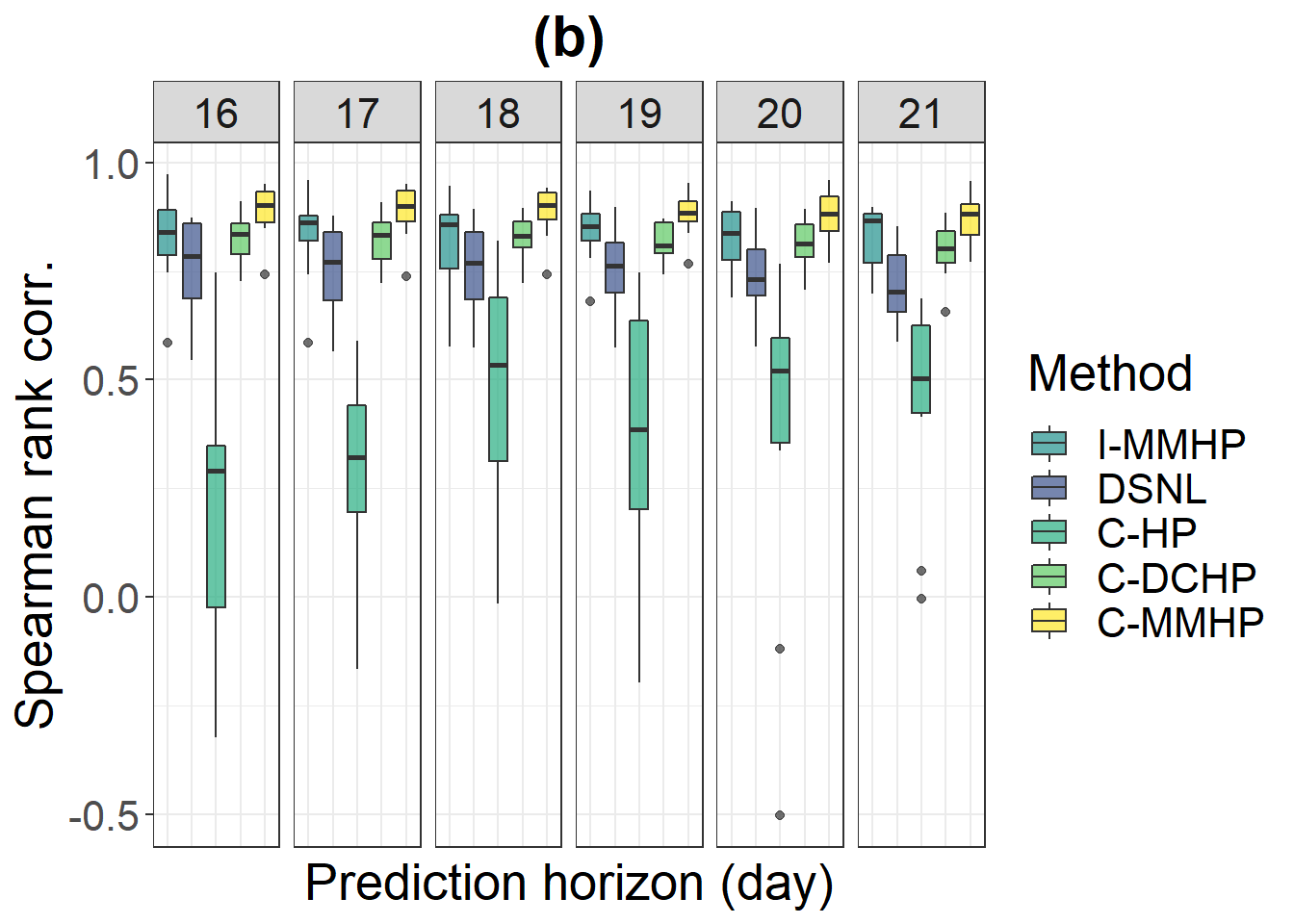}
 }
 \caption{Prediction of events and rank. (a) shows
 the MAE 
 of predicted error for all cohorts,
 using the median predicted count for each model for each cohort on 
 each day. (b) Summarizes the Spearman rank correlation of predicted 
 rank for all cohorts, where each cohort is predicted by 
 the posterior mean of $\hat{\lambda}_i(t^{(d)})$.}
 \label{fig_latent:real_predict_rank}
\end{figure}

Our posterior predictive processes can even be used to
forecast the Glicko scores over future prediction windows,
since we obtain the full event history from the generated process. In contrast, the DSNL 
model can only provide day-level predictions,
which we have evaluated previously. 
Figure~\ref{fig_latent:real_predict_separation}-(a)
shows the prediction of Glicko scores over days 19-21
when fitting the data in the first 18 days to the C-MMHP model. Our
prediction bands can forecast temporal trends of Glicko ratings in the real data 
and provide an appropriate representation of the uncertainty in these predictions,
which we illustrate in Figure~\ref{fig_latent:real_predict_separation}-(a). 
These prediction bands
correctly separate the rankings of most animals, particularly
the highly ranked nodes, and capture the groups of rankings that seem to
have formed for this cohort. 
Being able to predict rank evolution
over time is not possible using existing methods and this could be
of use in the experimental design of these studies to guide
data collection, such as the ability
to isolate mice who would be expected to rank similarly in the original
group.

\subsubsection{Additional insights from the C-MMHP model}
Finally, we wish to highlight some additional insights which are available 
after fitting our C-MMHP model.
Since our C-MMHP model can separate wins into active and 
inactive states, such separation can serve as a prepossessing step for the data. 
To illustrate this,
we first fit the C-MMHP model to the data for one cohort and classify the 
wins into active and inactive states
according to the estimated latent 
Markov process. The two types of interactions can then be fitted separately 
using other animal behavior models.
\cite{wu_markov-modulated_2021} shows that the wins 
in the active state more closely follow a linear hierarchy, 
as compared to the set of all wins
or the set of inactive wins; this 
provides an explanation for the \textit{pair-flips} phenomenon.
During the active state, pairs are engaging in aggressive interactions and actively trying 
to navigate the social hierarchy,
while in the inactive state, the wins are more or
less random and lack specifically directed aggression seen in the active state. As an 
example, we fit the DSNL model to the set of overall events, 
active events and inactive events separately, and calculate the Spearman rank correlation 
between the latent ranks for each day as estimated by the DSNL model and the Glicko ratings
at the end of each day. Figure~\ref{fig_latent:real_predict_separation}-(b) shows these 
rank correlations on each day for the three types of wins.
This suggests that the 
information contained in these wins
varies over time, with the inactive state
showing a stronger correlation with the Glicko ranking in
the start and middle of the observation 
period.
Similarly, as \cite{williamson2016temporal} pointed out, the distribution of
individual dominance power within animal groups is an important question.
One way we can
address this is by looking at the ``out" degree estimates from our C-MMHP model.
In particular, we see significant agreement between the out degree parameter
estimates and whether \cite{williamson2016temporal} could identify the dominant 
animal after the first week. Where this animal could be clearly identified early on,
the C-MMHP model
resulted in a much larger out degree estimate for that animal than the
rest of that cohort. This was not the case for the two cohorts where the dominant
animal took longer to establish itself. This out degree parameter may therefore 
provide some insights into the distribution of individual dominance power within 
a given group, although further work is needed to further investigate this
important question posed by \cite{williamson2016temporal}.
Interestingly, a similar connection 
is not seen with the less flexible C-DCHP model, which
does not first classify events as active
and inactive and base the inferred ranking
on only the active events.


\begin{figure}[ht]
 \subfigure
 {
\includegraphics[width=0.47\linewidth]{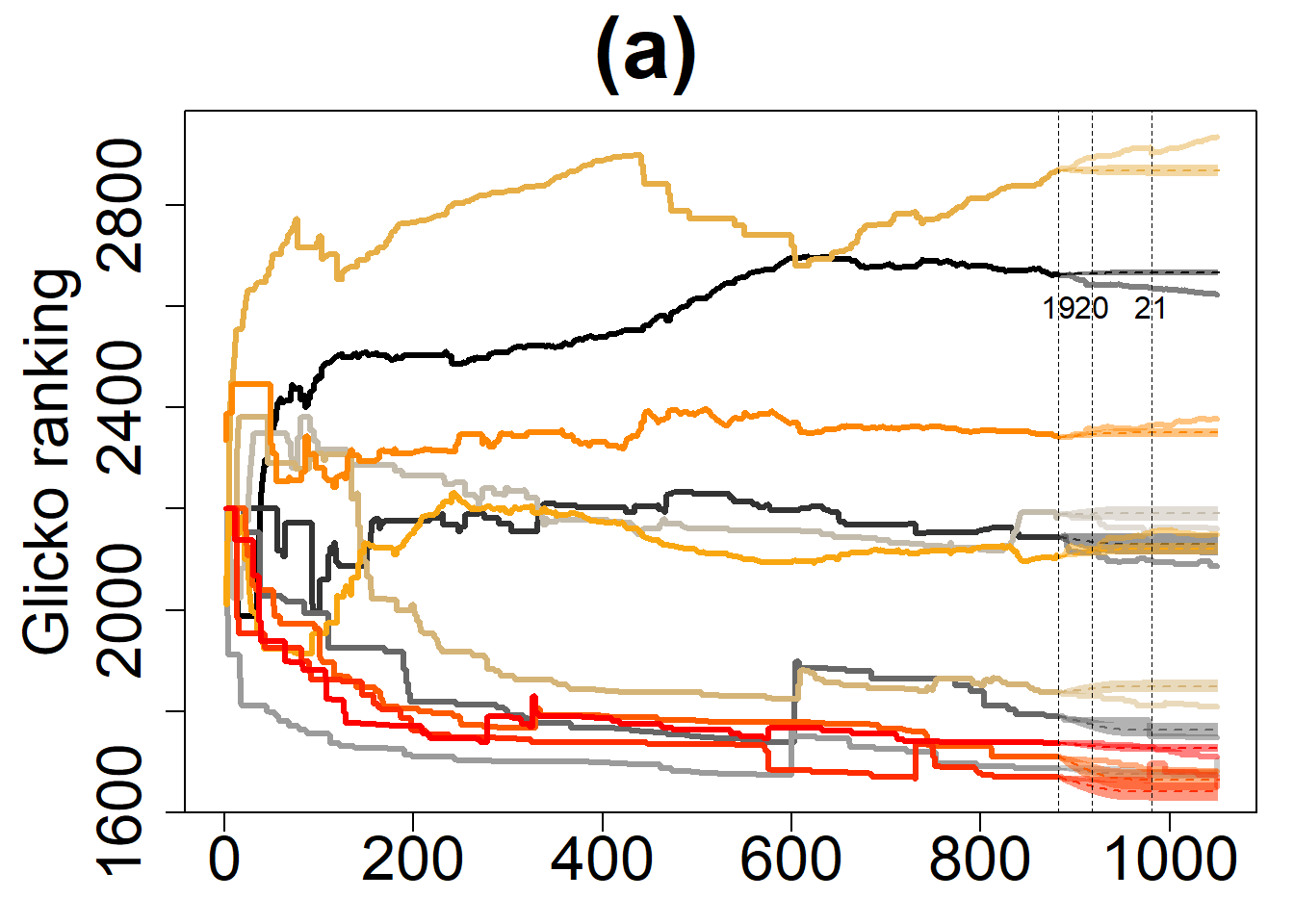}
 }
 \subfigure
 {
\includegraphics[width=0.47\linewidth]{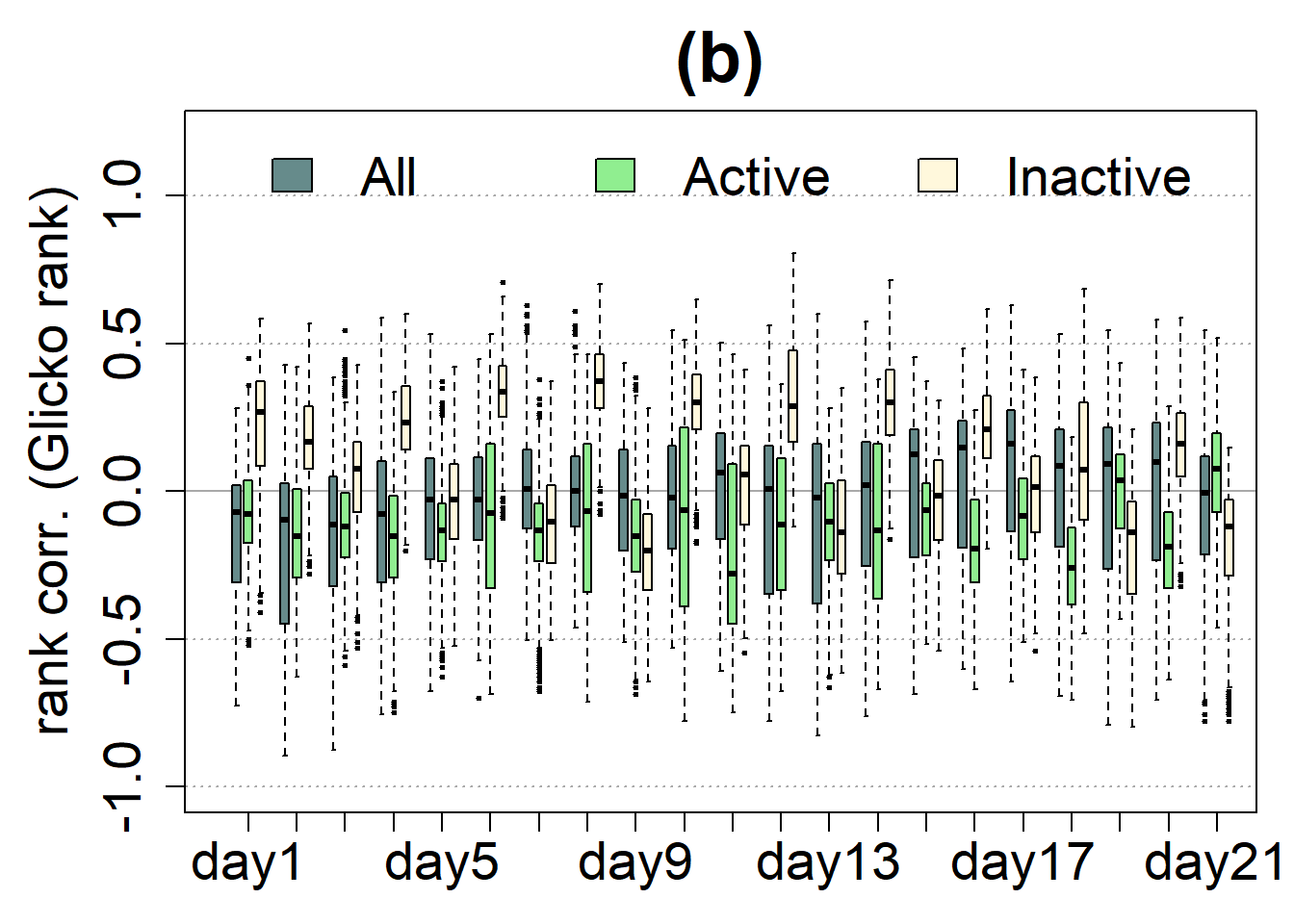}
 }
 \caption{Further results on C-MMHP. (a) Glicko score ranking prediction 
 of last three days using posterior draws, after fitting the first 18 
 days data in C-MMHP. True Glicko score ranking of all the time period is
 shown with the
 solid colored line, while the posterior prediction mean is in dashed 
 line and one standard deviation is plotted in shaded color. 
 The x-axis corresponds to the total number of interactions 
 across the cohort.
 (b) Rank 
 correlation between DSNL inferred latent rank and Glicko score ranking
 for each day in one cohort.
 Three colored bar indicated the performance of
 three inferred rankings conducted on the overall interactions, active 
 and inactive respectively.}
 \label{fig_latent:real_predict_separation}
\end{figure}





\section{Discussion}
\label{sec_latent:discussion}
In this paper, we propose a series statistical models 
that can uncover latent social 
dominance hierarchy among a group of animals from 
interaction win event times.
These models can serve as an important tool in animal aggressive behavior 
analysis and provide insight into important questions 
of recognition and its role in hierarchy formation.
To accomplish this, we formalize a point process model for 
continuous-time directed social network data. Three such 
models are developed:
the 
cohort Hawkes process model (C-HP),
the cohort degree-corrected Hawkes 
process 
model (C-DCHP) and the cohort 
Markov-modulated Hawkes process model (C-MMHP). 
The Hawkes process incorporates
the winner effect and bursting patterns of 
aggressive behaviors, 
which are regularly observed in patterns of aggressive interactions
across animal species. The degree correction
allows the model to better capture 
individual level heterogeneity that
is commonly observed in data of this form.
Finally, Markov-modulation accounts for
pair-flip situations and
allows for asymmetry in
interactions between pairs of animals,
by separating these
interactions into active and inactive states.
Performing inference for these models in the Bayesian
paradigm allows us to accurately quantify the uncertainty
in the inferred rankings and to 
better infer the ranking of nodes involved in 
few interactions, components that have been lacking 
in existing models for animal ranking.

The simulation study demonstrates
that inferences from these models are reasonable
and that the true ranking of nodes can be recovered.
The mice cohort study serves 
as a real data example and demonstrates that
the C-MMHP model performs best 
overall,
in terms of
providing insightful
latent rank inference, prediction
of both events and rank over time and potentially being
of interest in generating future research directions.
Although we do not have a ground truth for rankings in real data,
we have described how our model complements existing
ranking methods in the literature and the potential
value of the additional inference available. 
That the event dynamics can be described by a latent ranking
provides evidence that how these mice interact and explore their
social structure is driven by their position in a hierarchy, a key idea
in recognition.

We explore the dynamics of animal behavior using this model
motivated by observed and hypothesized phenomena 
in data of this form.
The also highlight how the use of a new model
may capture phenomena which existing models are not suited for.
The results from our analysis of aggressive
mice interactions provide insights on the agreement between 
model assumptions and the observed dynamics.
This C-MMHP model can also be used to simulated future events,
which could aid in designing studies of this form. Similarly,
the state separation available in the C-MMHP model
could lead to additional insights in conjunction with 
other models for animal behavior.

In the future, our model can be extended to incorporate the loser effect 
and bystander effect \citep{chase2011self} within the Hawkes process 
intensity function. The loser effect means that an animal 
that has lost 
in earlier contests has an increased probability of losing subsequent 
contests with other individuals. The bystander effect describes the 
situation where an animal's behavior might be influenced by observing an 
interaction or contest between two other animals. The extent of each 
effect can be estimated through a multivariate Hawkes process. The 
existence of such effects could be tested through the limiting 
distribution in \cite{chen2017multivariate}. 
Alternatively, models which utilise reciprocating interactions,
such as \cite{blundell2012modelling} could also be
of interest.
\cite{so2015social} raises a
question about the causal relationship between aggressive behavior and 
gene expression. It is feasible to integrate these elements in our model 
by modeling the baseline intensities as a function of covariates that 
correspond to gene expression. 
We have also seen that certain global
parameters in our models do not vary from Cohort to Cohort. As 
such, it would be of interest to design a hierarchical version
of our model, borrowing strength from different datasets.
Further work could also be done to make the degree estimates
a function of the nodes latent rank, although there is no 
clear evidence from the literature as to the 
association between the baseline activity level and
the position in a realised hierarchy. This question therefor remains
an important future problem.

Similarly, the model
we have proposed here is a special case of a latent space model.
Latent space models are an important tool in social network analysis
and have been widely used in modeling both static network 
\citep{de2018physical, hoff2005bilinear, mccormick2015latent} and dynamic network 
\citep{sarkar2006dynamic,sewell2015latent} data.
Although latent space models
of discrete-time dynamic networks have been considered \citep{kim2018review},
along with continuous time dynamics across repeatedly observed matrices
\citep{durante2014nonparametric},
there has been little work in the context
of continuous time events occurring on networks, and this remains an area
for future research.


\section*{Data and Code}
All data and code used in this paper are available
in a public repository.


\bibliographystyle{apalike}
\bibliography{ref_latent}
\end{document}